\documentclass[twocolappendix,numberedappendix]{emulateapj}
\usepackage{amsmath}
\usepackage{float}   
\usepackage{color}
\usepackage{rotating}
\usepackage{placeins} 
\usepackage{comment}
\usepackage{graphicx}
\usepackage{epstopdf}
\usepackage{verbatim}

\def\etal{{et~al.}}

\slugcomment{Accepted for Publication in the Astrophysical Journal}
\shorttitle{The Clustering of Young Stellar Clusters in NGC 628}
\shortauthors{Grasha \etal}

\begin{document}

\title{The Spatial Distribution of the Young Stellar Clusters in the Star Forming Galaxy NGC 628}
\author{K. Grasha\altaffilmark{1}, 
D. Calzetti\altaffilmark{1}, 
A. Adamo\altaffilmark{2}, 
H. Kim\altaffilmark{3,4}, 
B.G. Elmegreen\altaffilmark{5}, 
D.A. Gouliermis\altaffilmark{6,7}, 
A. Aloisi\altaffilmark{8}, 
S.N. Bright\altaffilmark{8}, 
C. Christian \altaffilmark{8}, 
M. Cignoni\altaffilmark{8}, 
D.A. Dale\altaffilmark{9}, 
C. Dobbs\altaffilmark{10}, 
D.M. Elmegreen\altaffilmark{11}, 
M. Fumagalli\altaffilmark{12}, 
J.S. Gallagher III\altaffilmark{13}, 
E.K. Grebel\altaffilmark{14}, 
K.E. Johnson\altaffilmark{15}, 
J.C. Lee\altaffilmark{8,19}
M. Messa\altaffilmark{2}, 
L.J. Smith\altaffilmark{16}, 
J.E. Ryon\altaffilmark{13}, 
D. Thilker\altaffilmark{17}, 
L. Ubeda\altaffilmark{8},
A. Wofford\altaffilmark{18}}
\altaffiltext{1}{Astronomy Department, University of Massachusetts, Amherst, MA 01003, USA; kgrasha@astro.umass.edu}
\altaffiltext{2}{Dept. of Astronomy, The Oskar Klein Centre, Stockholm University, Stockholm, Sweden}
\altaffiltext{3}{Korea Astronomy and Space Science Institute, Daejeon, Republic of Korea}
\altaffiltext{4}{Dept. of Astronomy, University of Texas at Austin, Austin, TX}
\altaffiltext{5}{IBM Research Division, T.J. Watson Research Center, Yorktown Hts., NY}
\altaffiltext{6}{University of Heidelberg, Centre for Astronomy, Institute for Theoretical Astrophysics, Albert-Ueberle-Str.\,2, 69120 Heidelberg, Germany}
\altaffiltext{7}{Max Planck Institute for Astronomy,  K\"{o}nigstuhl\,17, 69117 Heidelberg, Germany}
\altaffiltext{8}{Space Telescope Science Institute, Baltimore, MD}
\altaffiltext{9}{Dept. of Physics and Astronomy, University of Wyoming, Laramie, WY}
\altaffiltext{10}{School of Physics and Astronomy, University of Exeter, Exeter, United Kingdom}
\altaffiltext{11}{Dept. of Physics and Astronomy, Vassar College, Poughkeepsie, NY}
\altaffiltext{12}{Institute for Computational Cosmology and Centre for Extragalactic Astronomy, Department of Physics, Durham University, Durham, United Kingdom}
\altaffiltext{13}{Dept. of Astronomy, University of Wisconsin--Madison, Madison, WI}
\altaffiltext{14}{Astronomisches Rechen-Institut, Zentrum f\"ur Astronomie der Universit\"at Heidelberg, Heidelberg, Germany}
\altaffiltext{15}{Dept. of Astronomy, University of Virginia, Charlottesville, VA}
\altaffiltext{16}{European Space Agency/Space Telescope Science Institute, Baltimore, MD}
\altaffiltext{17}{Dept. of Physics and Astronomy, The Johns Hopkins University, Baltimore, MD}
\altaffiltext{18}{UPMC--CNRS, UMR7095, Institut d'Astrophysique de Paris, Paris, France}
\altaffiltext{19}{Visiting Astronomer, Spitzer Science Center, Caltech. Pasadena, CA}

\begin{abstract}
We present a study of the spatial distribution of the stellar cluster populations in the star forming galaxy NGC~628.  Using Hubble Space Telescope broad band WFC3/UVIS UV and optical images from the Treasury Program LEGUS (Legacy ExtraGalactic UV Survey), we have identified 1392 potential young ($\lesssim 100$~Myr) stellar clusters within the galaxy, identified from a combination of visual inspection and automatic selection.   We investigate the clustering of these young stellar clusters and quantify the strength and change of clustering strength with scale using the two-point correlation function.  We also investigate how image boundary conditions and dust lanes affect the observed clustering.  The distribution of the clusters is well fit by a broken power law with negative exponent $\alpha$.  We recover a weighted mean index of $\alpha \sim -0.8$ for all spatial scales below the break at 3\farcs3 (158~pc at a distance of 9.9~Mpc) and an index of $\alpha \sim -0.18$ above 158 pc for the accumulation of all cluster types.  The strength of the clustering increases with decreasing age and clusters older than 40 Myr lose their clustered structure very rapidly and tend to be randomly distributed in this galaxy whereas the mass of the star cluster has little effect on the clustering strength.  This is consistent with results from other studies that the morphological hierarchy in stellar clustering resembles the same hierarchy as the turbulent interstellar medium.  
\end{abstract}
\keywords{galaxies: star clusters -- galaxies: star formation -- ultraviolet: galaxies -- galaxies: individual: NGC 628}



\section{Introduction}\label{sec:intro}
Star formation leads to the creation of stellar clusters \citep{lada03} and most, if not all, stars form in some type of clustered structure.  Observations of local star forming regions have shown that clustering is a common feature, resulting from the fractal properties of the interstellar medium (ISM) under the effects of turbulence \citep{elmegreen97,elmegreen14}.  As a result, gravitationally bound clusters occupy the smallest and densest regions of the hierarchy of giant molecular cloud (GMC) complexes forming on the larger scales ($\sim$1 kpc).  

As clusters can be observed to greater distances than individual stars, young stellar clusters provide an excellent means to investigate the connection between the continuous distribution of star formation on small scales to galactic formation at large scales.  While studies on the hierarchical clustering of stellar populations have become more sophisticated over time \citep[e.g., see the early papers of ][]{payne74,efremov95}, whether or not the stars and the clusters are mapping the same type of hierarchy (i.e., the dense peaks are randomly distributed within each hierarchy or they are a biased representation of each hierarchy) is still a question that needs to be answered.  The answer will help understand the role of cluster formation and evolution that is complementary to studies of the cluster formation efficiency.  The spatial distribution of newly formed stars is also important as it provides insight not only on the processes of star formation, but also on the evolution and environmental dependences within stellar clusters and their migration from their clustered complexes.  For instance, we expect that the observed hierarchical clustering disappear with age \citep{elmegreen06,elmegreen10}: the densest regions with the shortest mixing timescales lose their substructures first, whereas the larger, unbound regions will lose their substructure over time owing to random initial motions and tidal forces.  The migration timescale for which stars and clusters `abandon' their clustered natal structure is not well constrained for most systems, but such knowledge is vital toward understanding how star formation evolves in both space and time.  

One of the most powerful ways to probe the clustering distribution of galactic components is with the two-point correlation function \citep{peebles80,zhang01,odekon06,odekon08}, a statistical tool to quantify the excess probability of finding one object within a specified distance of another object against that of a random, unclustered distribution.  Applying correlation functions to study the clustering of clusters will provide insight to the (1) physical process of cluster formation; (2) the extent that formation of clusters is hierarchical; and (3) whether or not the clustered distribution of young stellar clusters reflects the fractal structure of the interstellar gas \citep{efremov98,elmegreen96,elmegreenelmegreen01,bastian07}.  

Few studies have been done on the clustering distribution of stellar clusters thus far, showing the need for a systematic approach to the problem across a wide variety of galaxies and environments.  In one of the first studies on the hierarchy occurring within stellar cluster ensembles, \citet{zhang01} found that a power-law correlation function well described the clustering hierarchy up to 1~kpc for the star clusters within the Antennae galaxies.  \citet{scheepmaker09}, in a study of the stellar clusters in M~51, found that clusters showed an age-dependency in their degree of clustering, where the youngest clusters were more clustered compared to older clusters.  

Investigations of clustering behavior across time and scale lengths on stellar systems have been performed in a few local galaxies in recent years as well.  In a study of the stellar samples of M 31, \citet{gouliermis15a} found that the youngest ($<$25~Myr) stars were more clustered than older stars ($<$300~Myr) and that the observed clustering changed within different regions of the galaxy.  Studies of the stellar structures in both the Large Magellanic Cloud (LMC) and Small Magellanic Cloud (SMC) have shown that stars are born highly fractal and evolve toward a uniform distribution within the crossing time of the galaxy \citep[$\sim 175$ and 80 Myr for the LMC and SMC, respectively; ][]{gieles08,bastian09,bonatto10}.  Within NGC~1313, \citet{pellerin07} found that infant mortality --  the process by which most clusters do not survive the death of the most massive stars in the cluster, forcing the remaining unbound clusters to dissolve over a short period of time \citep[$\lesssim 10$~Myr: ][]{lada03,fall05,bastian06,chandar10b} --  is an efficient process in disrupting, and therefore, destroying stellar clusters by showing that B-stars can be found outside of star clusters.  These unbound stars in the diffuse field can account for a large portion of the UV light from a galaxy \citep{meurer95,tremonti01,hoopes01,crocker15} and the disruption and dissolution of clusters is crucial in populating the diffuse stellar field.  The evolution of the clustering of star clusters with time will thus depend on the combination of two factors: how fast the cluster migrates within a galaxy and how long the typical star cluster survives.  

If the cluster dispersion timescale has a dependency with galactic environment, we would expect to see their age distribution change over different ambient environments within a galaxy.  Indeed, in a study by \citet{sanchezgil11}, the young stellar populations within NGC~628 exhibited age gradients across the spiral arms, with the youngest star formation regions concentrated along the center of the spirals, and an age gradient from the inner to the outer parts of the galaxy.  Radial stellar migration can further flatten the observed age gradient, implying steeper age gradients than those measured.  Additionally, \citet{bastian11a} have found strong dependency of clusters age with galaxy position, where the age distribution varies as a function of galactocentric distance, becoming shallower toward the outskirts of the galaxy.  Theoretical work \citep{elmegreenhunter10, kruijssen11} suggests that the age distribution of clusters do vary with environment and that it can influence the dissolution timescale, with clusters disrupting faster when the gas surface density $\Sigma_{\rm gas}$ is high and living longer when $\Sigma_{\rm gas}$ is low.  On the other hand, if cluster dispersion is independent of environment, the age distribution of all clusters should be the same, regardless of position within a galaxy.  Studies of both LMC and SMC have shown that there is a flat age distribution of clusters, concluding that cluster disruption does not significantly affect the age distribution and that there is little dependence on environment for the first $\sim$200 Myr  \citep{gieles07,baumgardt13}.  Early results of the Panchromatic Hubble Andromeda Treasury \citep[PHAT;][]{dalcanton12} survey show that the clusters of M 31 show no evidence for cluster dissolution at early times (30--100 Myr).  Using the first PHAT clusters catalog of 601 clusters, \citet{fouesneau14} confirms that cluster disruption has little to no effect prior to the timescale of 100 Myr in Andromeda.  The spatial correlation between clusters will allow us to study the extent of evolution in the clusters and whether or not the formation of clusters is hierarchical \citep{efremov98,bastian07}.  

Ambient environment is crucial for how clusters form, and quite possibly, it is even more critical to investigate if stellar clusters trace star formation.  Studies on the clustering of stars have revealed important information on the nature of the star formation process itself.  The recent results of \citet{gouliermis15a} show an environmental dependence in the clustering distribution of stars in the nearby spiral M~33, where stars in the outer regions of the galaxy experienced less disruption and larger amounts of clustering compared to stars located in the inner regions.  This agrees with the work of \citet{silvavilla14}, which showed an environmental dependence on the cluster population within M~83, where the clusters in the outer region showed less disruption and flatter age distributions compared to clusters in the inner region.  Work by \citet{gouliermis15b} for NGC 6503 shows that younger stars follow a hierarchical distribution whereas older stars display a homogeneous, less clustered distribution.  By comparing the clustering results within this work to the clustered results of stars, we will be able to infer if and how star clusters also trace star formation.  

The nearby spiral galaxy in this study, NGC 628, was observed as part of the Legacy ExtraGalactic UV Survey\footnote{https://legus.stsci.edu/} \citep[LEGUS;][]{calzetti15}, a Cycle 21 Hubble Space Telescope (HST) program which has imaged 50 nearby ($\sim$3.5--12~Mpc) galaxies in five UV and optical bands (NUV,U,B,V,I) with the UV/Visible (UVIS) channel on the Wide Field Camera 3 (WFC3).  The aim of LEGUS is to investigate the relation between star formation and its galactic environment in nearby galaxies, over scales ranging from individual star systems to kpc-sized structures.  NGC 628 is a face on grand design spiral galaxy with a large number of star clusters to provide a statistically powerful test for changes in clustering strength with scale.  The occurrence of hierarchical structures within NGC 628 was already investigated by \citet{elmegreen06} using ACS data in B, V, I, and H$\alpha$ bands.  These authors examined the distributions of size and luminosity of star-forming regions, finding that both of these were well described with a power law, indicative of a hierarchical structure of stellar components within the galaxy.   

The work herein primarily focuses on describing the two-point correlation function as a tool to study the clustering properties of the young stellar clusters.  For this goal, we use the galaxy NGC628 as an example.  We hope to address the following points once the clustering properties of a larger number of LEGUS galaxies have been studied:  (1) whether stars clusters are clustered and how strongly; (2) how the clustered distribution of stars compare to that of star clusters; (3) how the clustering depends on age and environment and the dissolution timescale of the clusters; and (4) whether or not clusters can be used to trace the structure of star formation.  

The galaxy selection and cluster identification process is described in Section \ref{sec:ngc628}.  The methodology of two-point correlation function is introduced in Section \ref{sec:2pcf}.  In Section \ref{sec:results}, we describe the results and analysis and how we use the correlation function to draw conclusions about the properties of our star clusters.  We discuss our results concerning hierarchy of the stellar clusters in Section \ref{sec:discussion}.  Finally, we summarize the findings of this study in Section \ref{sec:summary}.

\section{NGC 628}\label{sec:ngc628}
We study the face-on grand design spiral galaxy NGC 628 located at a distance of 9.9~Mpc \citep{olivares10} with no apparent bulge.  NGC 628 is the largest galaxy of the M74 galaxy group and has a global SFR(UV) of approximately 3.7 $M_{\odot}~{\rm yr}^{-1}$ \citep{lee09}.  The stellar disk appears largely undisturbed in studies of the gas kinematics with optical and UV imaging \citep{herbertfort10} and has a disk thickness of 0.25 kpc \citep{peng88}.  We select this galaxy primarily due to the low-inclination angle (25 degrees), relatively large angular size (10'.5 x 9'.5), clearly defined spiral arms, and high number of clusters available for analysis. 

\begin{figure*}
\epsscale{1.08}
\plottwo{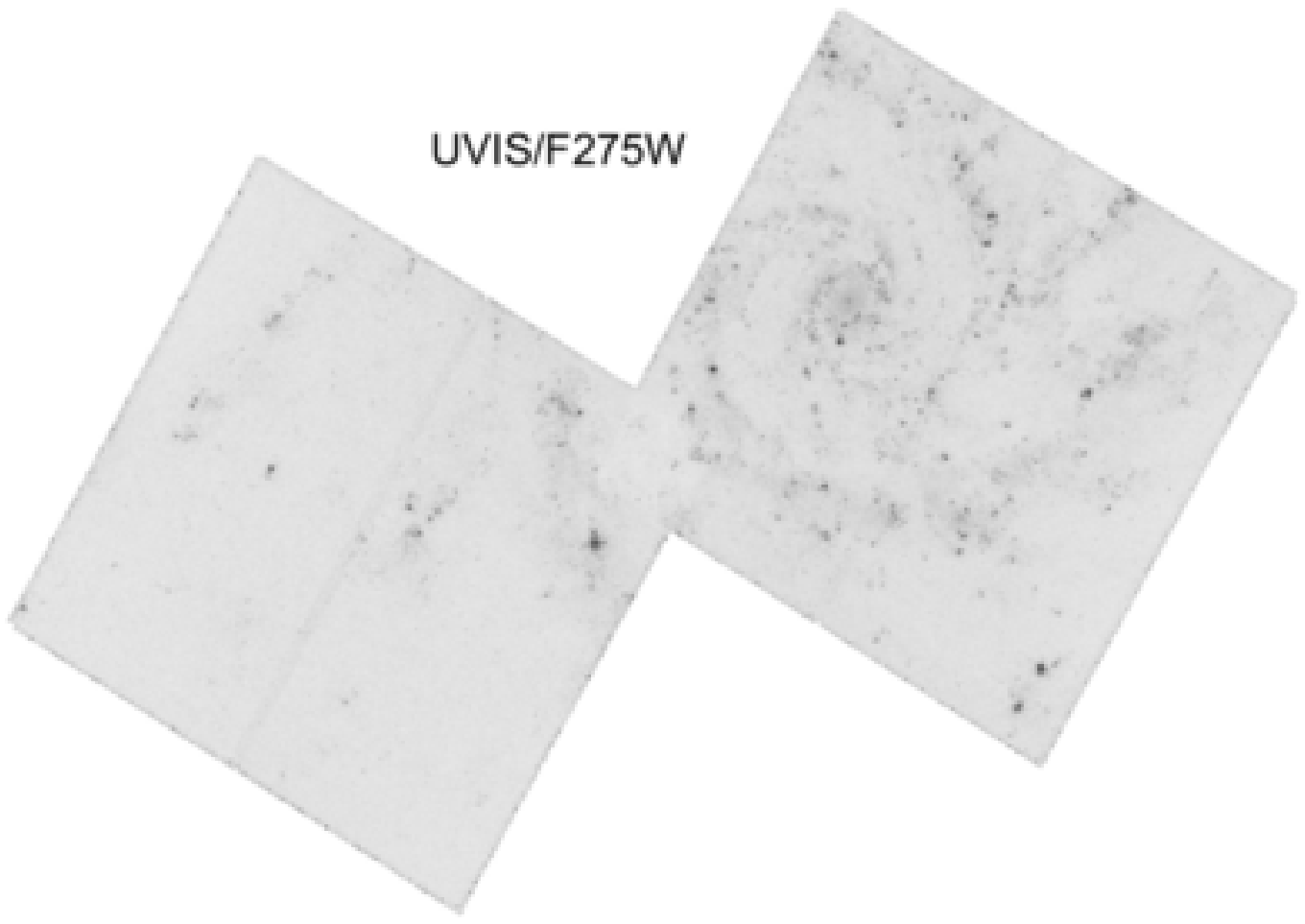}{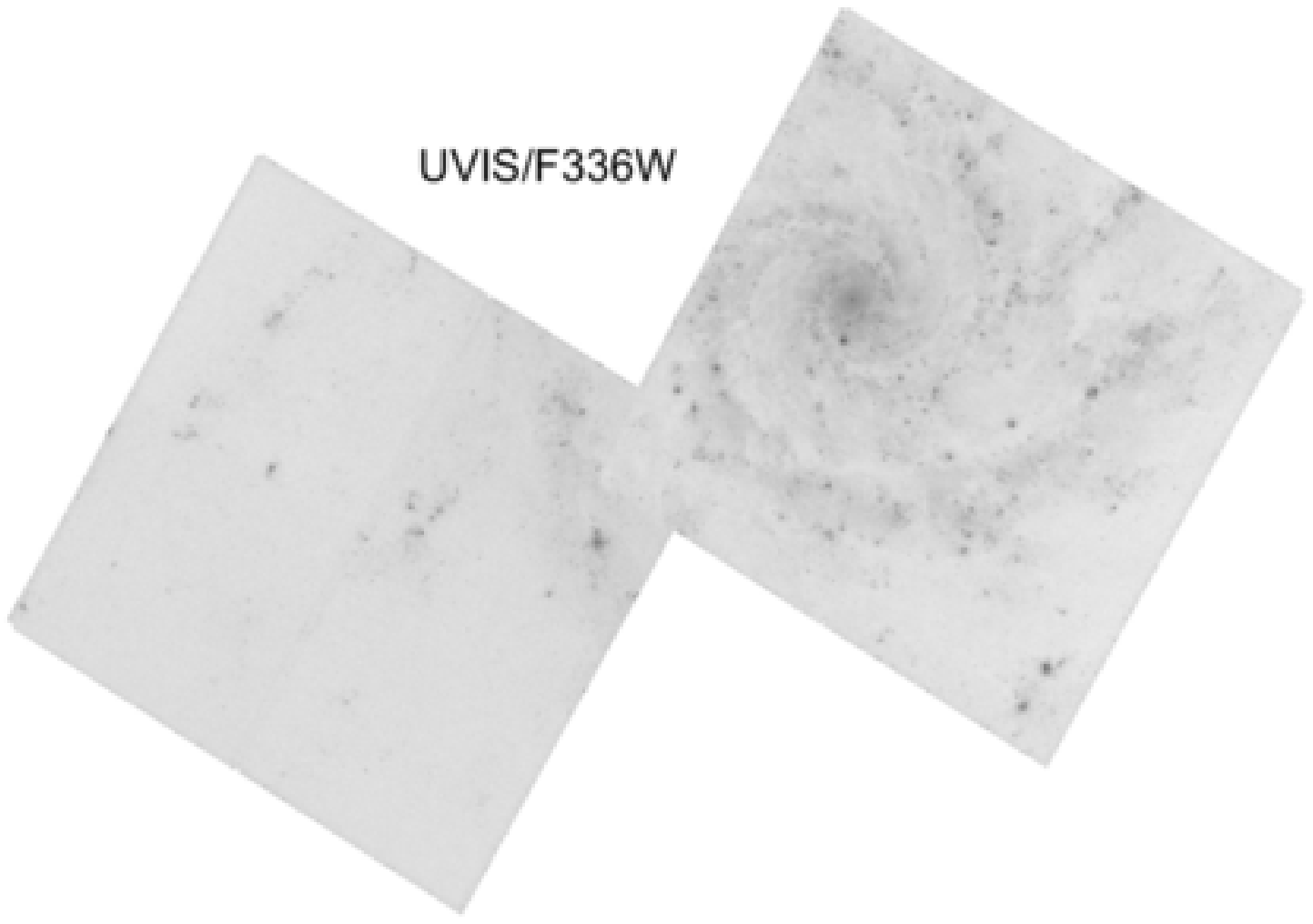}
\plottwo{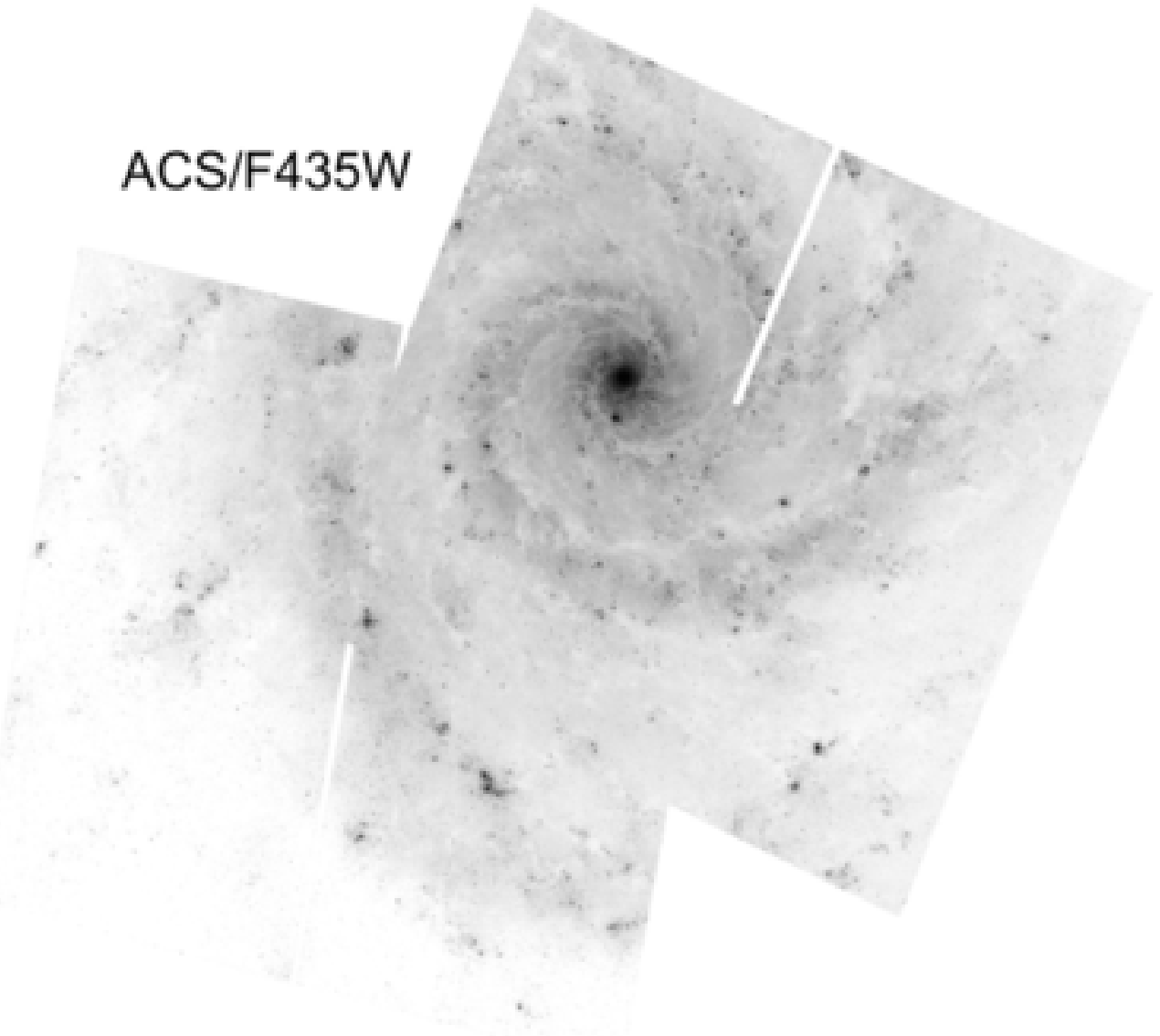}{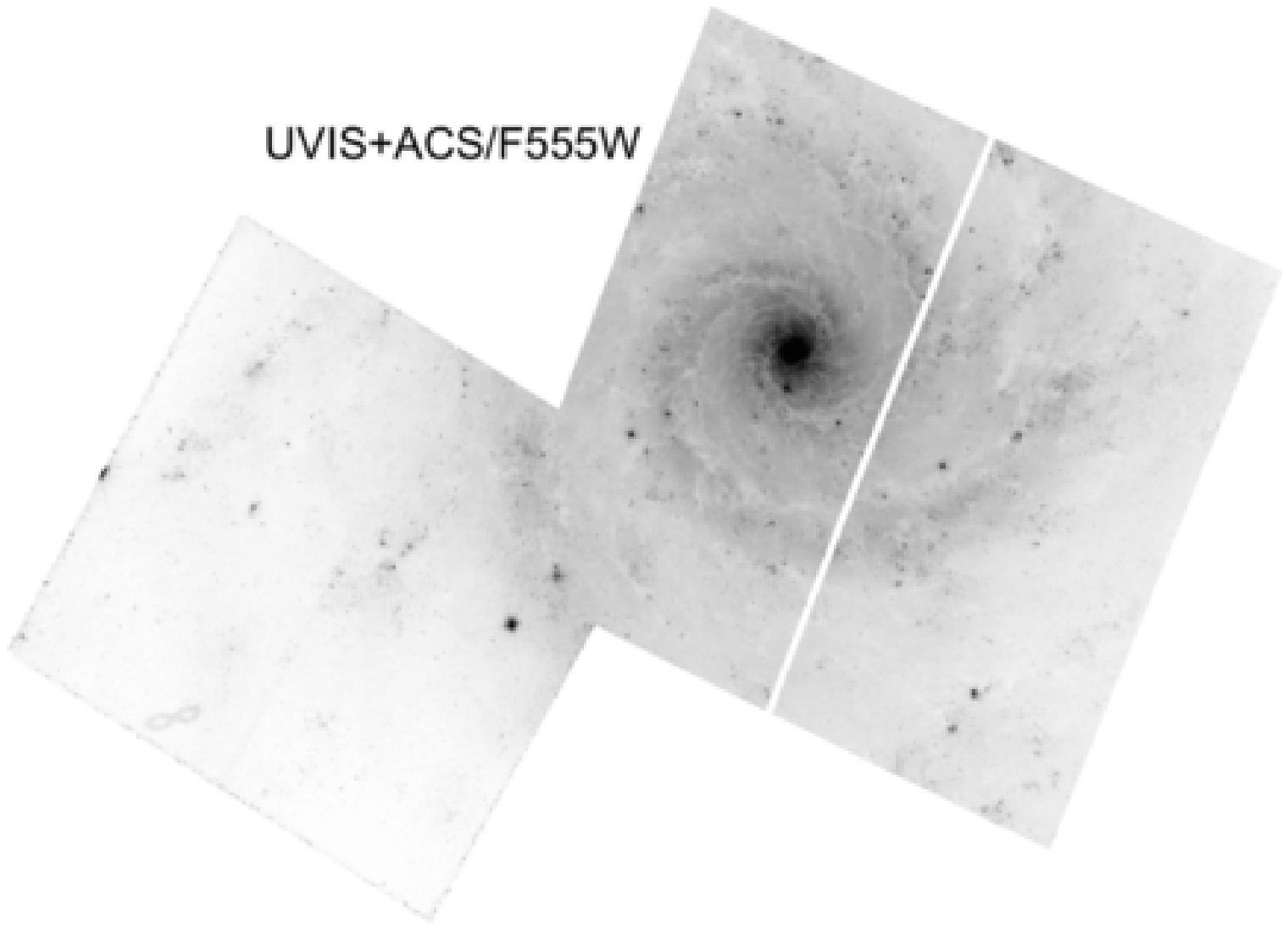}
\plottwo{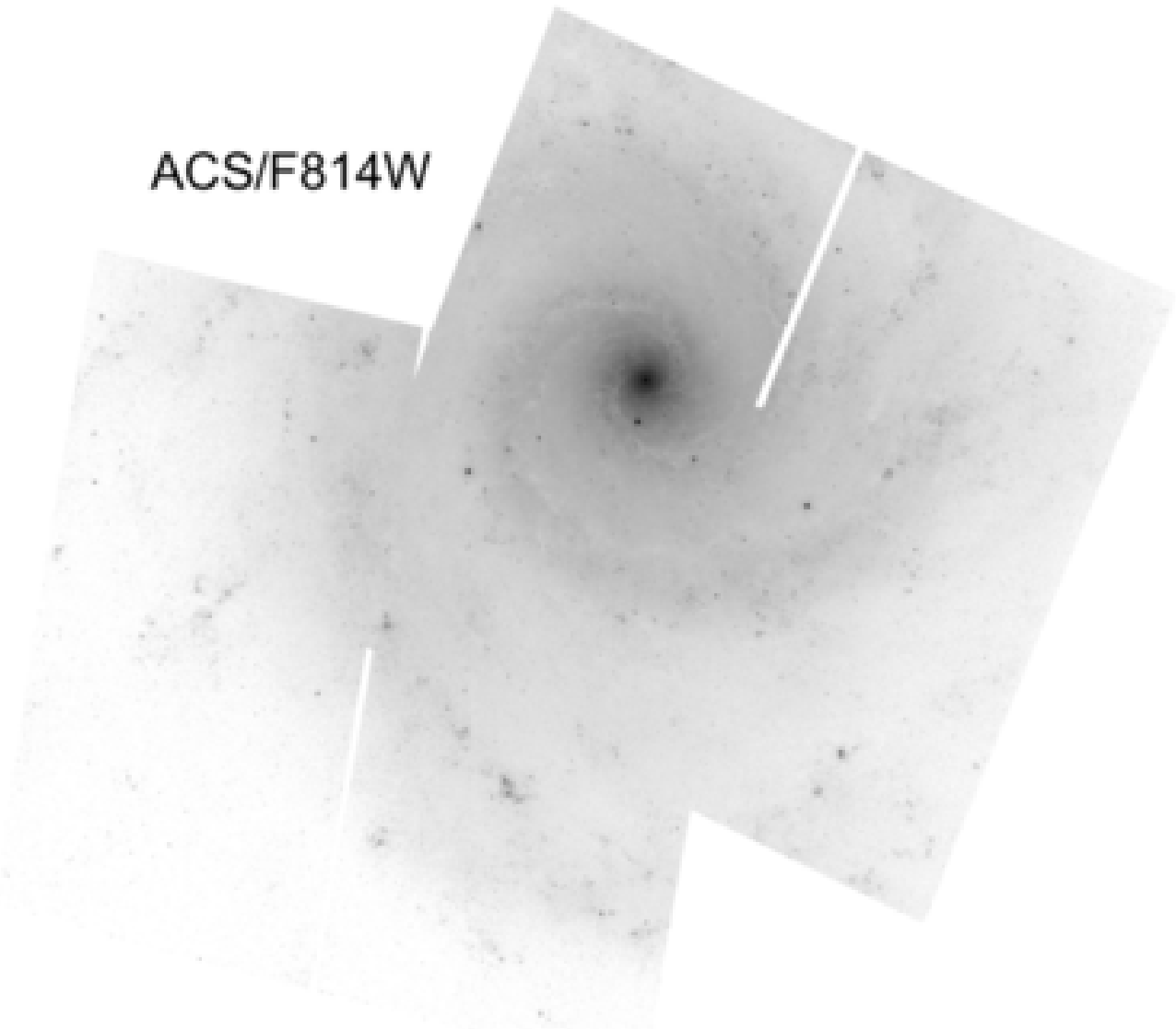}{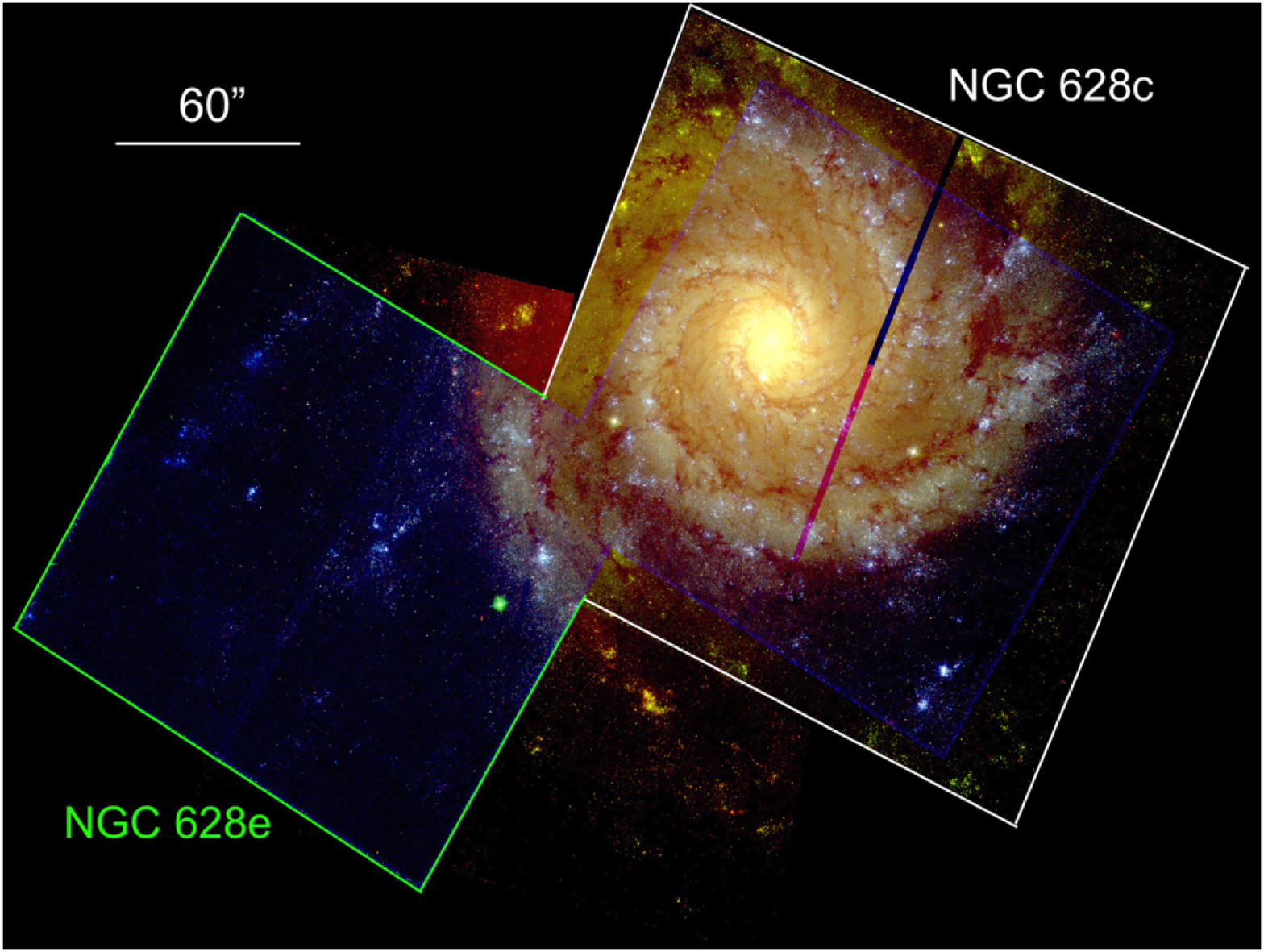}
\caption{
Gray-scale mosaic images of the five broad band filters of NGC 628 (image rotated with North up).  The bottom right shows the RGB color composite mosaic UVIS/F275W and UVIS/F336W (blue), UVIS/F438W and UVIS/F555W (green), and UVIS/F814W (red).  The white line outlines the central pointing, NGC 628c, and the green line outlines the east pointing, NGC 628e.  The white horizontal bar in the upper left denotes the length scale of 60'', 3.3 kpc at the distance of NGC 628.  
\label{fig:gal}}
\end{figure*}

The images of NGC 628 for our analysis were obtained in five broad band filters, NUV (F275W), U (F336W), B (F438W), V (F555W) and I (F814W) in October 2013 with the UV-optical channel (UVIS).  Observations of NGC 628 consist of two pointings, one towards the center of the galaxy, NGC628c, and one towards the galaxy's eastern region, NGC628e, adjacent to the first pointing.  All five filters in both pointings are aligned and referenced to the UVIS/F336W filter.  Both pointings are combined into a single mosaic for analysis, shown in Figure \ref{fig:gal}.  The images have a scale of 0.04 arcsec per pixel, corresponding to a pixel resolution scale of 1.9 pc at a distance of 9.9~Mpc.  General descriptions of the standard data reduction of the LEGUS datasets is available in \citet{calzetti15}.

\section{Cluster Selection and Identification}\label{sec:clusterselection}
Stellar clusters within NGC 628 are identified by first undergoing an automated process using SExtractor \citep{bertin96} that avoids background sources and single, bright stars.  Each catalog includes sources which satisfy the two following conditions: (1) the V band concentration index (CI) must be greater than the stellar CI peak value (CI=1.4 and 1.3 for the inner and outer pointing, respectively, in order to make the selection criteria more uniform between the varying pixel scales of the cameras); and (2) the source should be detected in two contiguous bands with a signal-to-noise greater than 3.  

As stars appear unresolved even at the highest HST resolution power, their CI will vary little and their CI distribution will be highly picked around an average value typical of a stellar PSF.  On the other hand, clusters are partially resolved and their sizes can vary, therefore, they generally have CI larger than that of stellar values.  Using CI distributions of the extracted cluster candidates, we selected the smallest CI value that would allow us to remove the bulk of the stellar interlopers from our catalog.  Since the resolution power is also dependent of the HST camera used, the values used for the inner and outer frame of NGC 628 are different.  Simulations of CI as function of effective radius of the clusters show us that the chosen limits corresponds to $R_{\rm eff}=1$~pc at the distance of NGC 628.  

In order to secure reliable measurements from SED fitting, each source is required to have a 3$\sigma$ detection in at least four of the five photometric bands, necessary to adequately break the age-extinction degeneracy.  The physical properties of each cluster (ages, extinctions, masses) have been derived using deterministic stellar population models \citep[Yggdrasil;][]{zackrisson11} and a $\chi^2$ fitting approach as described in \citet{calzetti15}, which includes uncertainties estimates \citep[see ][]{adamo10}.  In this paper we use standard catalogues which contains cluster physical properties derived with deterministic models with solar metallicity for both stars and gas, an average covering factor of 50\%, and a Milky Way extinction curve \citep{cardelli89} with a foreground E(B--V)=0.06.  Photometry is performed with a circular aperture of 4 pixels in radius, with the background measured within an annulus of 7 pixels in inner radius and 1 pixel in width.  Single stellar populations are used to determine the ages of our clusters; for clusters that are better described with multiple stellar populations, our reported age is recovering the mean age of the cluster.  

The uncertainties we derive in age, mass, and extinction are about 0.1 dex.  Within the LEGUS collaboration we are also undertaking cluster analysis based on stochastically sampled stellar libraries and Bayesian approaches \citep{krumholz15}.  A comparison between the standard approach based on Yggdrasil deterministic models \citep{zackrisson11} and the analysis performed with SLUG models \citep{dasilva12} shows that the average derived cluster properties do agree down to clusters masses of 5000 M$_{\odot}$.  Below this mass range we observe the largest uncertainties in derived ages, masses, and extinctions because, as already widely discussed in the literature \citep[e.g., ][]{cervino04}, deterministic models are not able to correctly interpret the large scatter in color caused by the stochastic sampling of the cluster IMF.  However, in this analysis, we are mainly focusing on a morphological division of the cluster catalogues, based on visual inspection.  We will also explore the change in clustering using two age ranges (i.e., 1 to 40 and larger than 40 Myr) and two mass ranges (below and above log(M)=3.6 M$_{\odot}$).  A realistic error to associate to the value of 40 Myr is the average uncertainty of 0.1 dex observed in deterministic cluster properties; this corresponds to $40 (+20/-10)$~Myr. The low mass range will be affected by stochastic effects and indeed we will discuss this point in Section 4.3.  More details will be given in a forthcoming paper \citep{adamo15a}.  

\begin{figure}
\epsscale{1.2}
\plotone{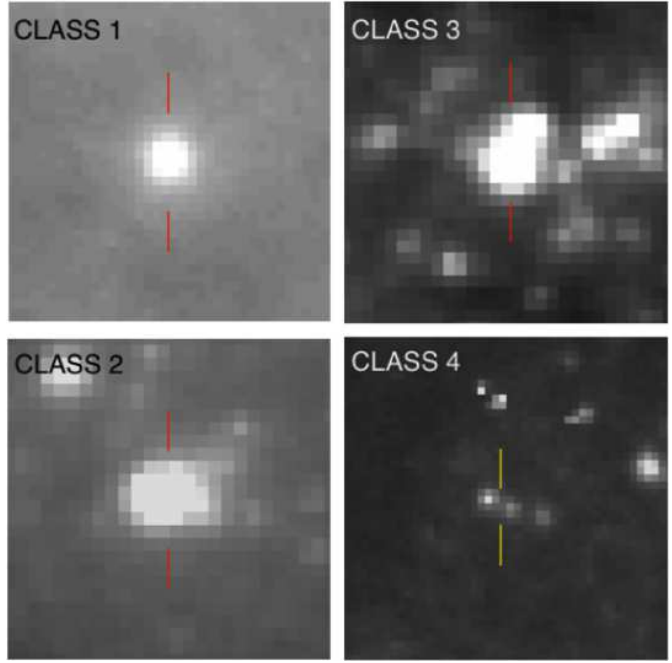}
\caption{
LEGUS classification system for clusters.  An example is shown for Class 1 (symmetrical), 2 (asymmetrical), 3 (multiple peak), and 4 (star or spurious object) in the F555W filter.  
\label{fig:classtypes}}
\end{figure}

Each object within the cluster catalog that has an absolute magnitude brighter than $-6$ mag (well within the 90\% completeness limits) in the V band is then visually inspected and assigned one of four categories:  (1) a symmetric, centrally concentrated cluster; (2) a concentrated cluster with some degree of asymmetry; (3) a multiple peaked system; or (4) a spurious detection such as a foreground/background source, single bright star, bad pixel, or a source that lies too close to the edge of the chip.  Class 4 objects are excluded from the final cluster catalog.  Figure \ref{fig:classtypes} shows what a typical cluster looks like for eac classification.  Classifications 1, 2, 3 are considered to be genuine star clusters or associations.  The classification of each object is compiled from visual inspection from at least four independent members of the LEGUS team.  The final cluster catalog is compiled by comparing all the results from each individual, after checking for consistency.  

\begin{figure}
\epsscale{1.2}
\plotone{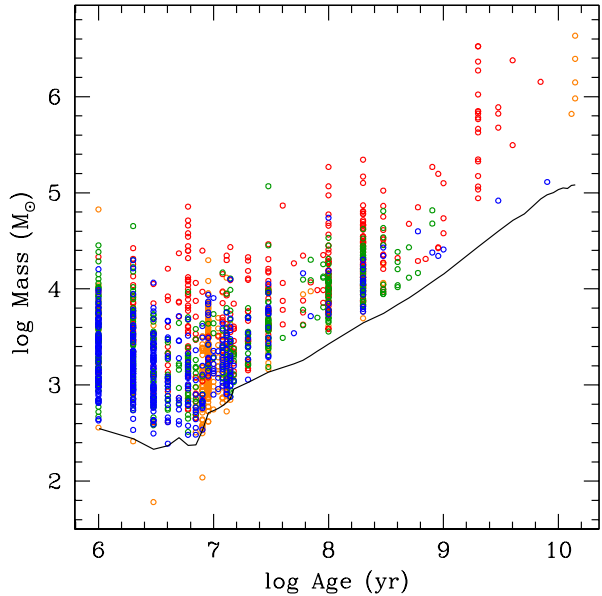}
\caption{
Age-mass diagram for star clusters in NGC 628.  The colors represent our different cluster classifications: Class 1 (symmetrical, red circles), 2 (asymmetrical, green circles), 3 (multiple peak, blue circles), and 0 (selected as fainter than  $M_{F555W} = -6.0$~mag, orange circles). The solid line is our 90\% completeness limit. 
\label{fig:agemass}}
\end{figure}

The distribution of masses and ages of the clusters is shown in Figure \ref{fig:agemass}, with the three classes identified by different colors.  In the same figure, the selection limit of $M_{F555W} = -6.0$~mag (apparent magnitude of 23.98 mag) for the clusters is shown by a continuous line, which follows the characteristic age-mass correlation for ages $>10^7$~yr.  General properties of the three cluster classes that will be detailed below include that class 3 clusters tend to be on average younger and less massive than Class 1 and Class 2 clusters. 

Cluster candidates that meet the requirements of a minimum cutoff of CI $>1.4$ (CI $>1.3$ for the clusters in NGC 628e) and detection in at least four filters but are fainter than an absolute V-band magnitude of $-6$ are not visually classified; however, these unclassified sources are still added to the cluster catalogs, labeled as Class 0 cluster candidates.  We are greatly interested in the faint cluster candidates in order to increase our sample, especially at the low-mass end \citep{krumholz15}.  Because these are not visually inspected sources, we acknowledge that there will be contamination of spurious objects in this class.  We impose additional constraints on the Class 0 (not visually inspected) objects by analyzing those with CI $\leq$ 1.6 mag separately from those with CI $>$ 1.6 mag.  This additional criterion of selecting the broadest of the Class 0 sources will increase our chance of identifying genuine clusters within the classification (see Section \ref{sec:agecuts} for a more detailed discussion).  For all Class 0 sources, we place a limit on the broadest CI of 4 to exclude background galaxies that may be included in the catalog.  We have identified 345 Class 0 cluster candidates.  For the 128 Class 0 candidates with CI $>$ 1.6 mag, we performed a visual check to quantify the accuracy of these objects being clusters.  The results can be found in Appendix \ref{sec:appA}, where we find that the contamination rate of non-clusters in the Class 0 is as high as 57\%.  

\begin{figure}
\epsscale{1.2}
\plotone{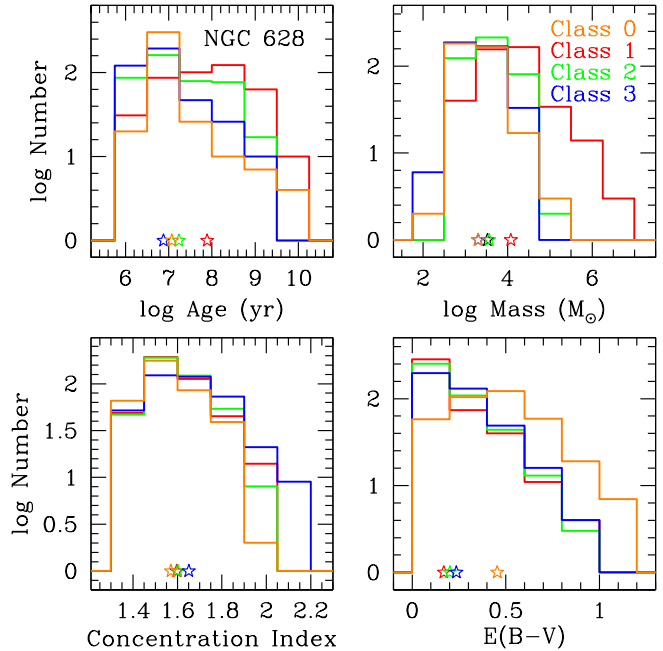}
\caption{
Histograms showing the break down of the clusters within NGC 628 as a function of age, mass, CI, and E(B--V), where the colors represent the cluster classification as defined in Section \ref{sec:clusterselection}:  Class 0 sources are orange, Class 1 are red, Class 2 are green, and Class 3 are blue.  The distribution of E(B--V) for Class 0 sources is notably different from the genuine star clusters.  The open star symbols at the bottom of each plot show the median value for each cluster classification. 
\label{fig:hist}}
\end{figure}

\begin{figure}
\epsscale{1.2}
\plotone{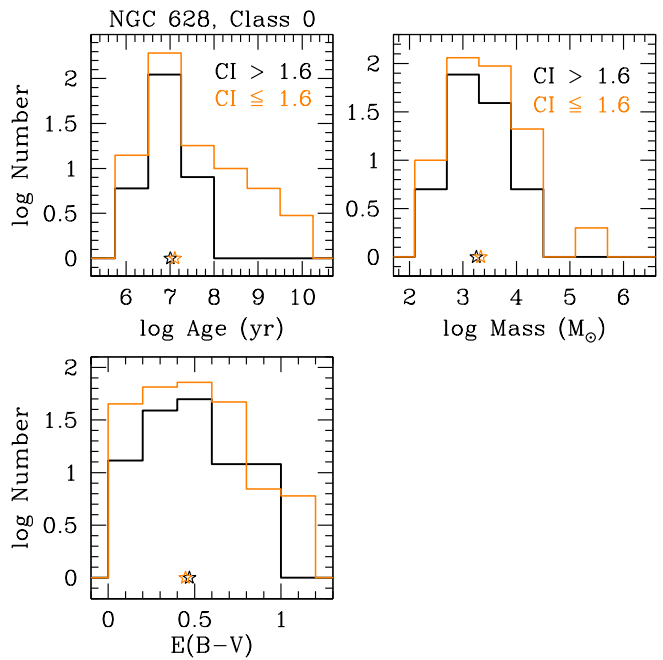}
\caption{
Histograms showing the break down of the Class 0 clusters within NGC 628 as a function of age, mass, and E(B--V) for the Class 0 clusters with a CI $>$ 1.6 mag (black solid line) and the clusters with CI $\leq$ 1.6 mag.  The open star symbols at the bottom of each plot show the median value for each cluster classification.  While the average value of mass or E(B--V) does not change between the sources with broader profiles compared to those with narrower profiles, there is an age difference in that CI$\leq$1.6 have a tail at larger ages. 
\label{fig:histclass0}}
\end{figure}

NGC 628c(e) has a total of 1019 (245) star clusters classified as Class 1, 2, or 3 and 345 (39) Class 0 cluster candidates with more than four filters available.  We combine both pointings for our analysis, resulting in a total of 1264 genuine (Class 1, 2, and 3) and 384 Class 0 cluster candidates.  There are 17 clusters in the overlapping region of the east and central pointings that are present in both catalogs; we have removed the duplicate clusters within the NGC 628e catalog.  The breakdown of all the classifications in age, mass, CI, and color excess E(B--V) are shown in Figure \ref{fig:hist}.  Figure \ref{fig:histclass0} shows the breakdown of only Class 0 objects with a CI cut at 1.6 mag, where we expect the sources with the broadest CI indices to exclude all non-cluster objects.  As expected, the clusters that are classified as multiple peak sources (Class 3) have the largest CI value on average.  Class 1 (centrally concentrated) clusters have larger masses and older ages when compared to the rest of the clusters in the galaxy.  The stellar clusters (Class 1, 2, 3) show small extinction, with an average color excess of E(B--V) $\sim 0.15$.  On the other hand, Class 0 candidates exhibit moderate extinction, with an average of E(B--V) = 0.46, which remains roughly the same for all Class 0 clusters, independent of CI.

\section{The Two-Point Correlation Function}\label{sec:2pcf}
Correlation functions provide statistical means by which to measure and provide detailed quantification on the clustering distributions of galaxy constituents.  One of the most commonly used quantitative measure of clustering structure is the three-dimensional two-point correlation function, $\xi(r)$, which measures the magnitude of clustering as a function of scale size.  

Originally defined by \citet{peebles80} as $\mathrm{d}P = \bar{n}^2 [ 1 + \xi(r) ]\ \mathrm{d}V_1 \mathrm{d}V_2$ for cosmological applications, the two-point correlation function $\xi(r)$ is defined as a measure of the probability of finding a neighboring object, above what is expected for an unclustered random Poisson distribution, in a shell element with volume element $\mathrm{d}V$ at a distance $r$ from any object with mean number density of the galactic structure, $\bar{n}$.  The two-point correlation function, $\xi(r)$, is usually fit as a power law, $\xi(r) = (r/r_0)^{-\gamma}$, where $r_0$ is the characteristic scale-length of the clustering, defined as the scale at which $\xi(r) = 1$.  The correlation function has been observed to have the same slope $\gamma$ across galaxy and cluster systems \citep[e.g., ][]{peebles75,postman98,daddi00,brodwin08}; the uniformity of the measured indices seem to suggest a common underlying dynamics on all scales.  

The spatial distribution of galactic components can also be measured in two-dimensions as projected onto the plane of the sky.  In our study, we measure the two-dimensional projected angular correlation function $\omega(\theta)$, defined as the probability above Poisson of finding two star clusters with an angular separation $\theta$ as $\mathrm{d}P = N^2 [ 1 +  \omega(\theta)]\ \mathrm{d}\Omega_1 \mathrm{d}\Omega_2$, where $N$ is the surface density of clusters per steradian with two infinitesimal elements of solid angle $\mathrm{d}\Omega_1$ and $\mathrm{d}\Omega_2$, separated by angle $\theta$.

To measure $\omega(\theta)$, pairs of stellar clusters are counted as a function of separation, compared by what is expected for an unclustered distribution.  A random catalog (in x, y position) of sources must be produced, populating the same sky coverage and geometry (e.g., edges, masks) as the data with randomly distributed points.  We define masks as areas that exclude all data, such as the ACS chip gap, or a reduction in the observed surface area of available data with respect to the global average, such as dust lanes or interarm regions of the galaxy.  We account for masks in the random catalog that are present in the real data.  The ratio of pairs of clusters observed in the data relative to pairs of points in the random catalog is then used to estimate $\omega(\theta)$.  In this study, we will implement the \citet[][LS]{landy93} estimator, which cross-correlates the data and random catalog to minimize edge and mask effects, written as, 
\begin{equation}\label{eq:LS}
\omega_{\rm LS}(\theta) = \frac{DD(\theta) - 2DR(\theta) + RR(\theta)}{RR(\theta)},
\end{equation}
where $DD$ is the number of data-data pairs, $DR$ is the number of cross-correlated data-random pairs, and $RR$ is the number of random-random pairs with the same mean density and sampling geometry with separation between $\theta$ and $\theta + \delta \theta$.  The pairs are computed as,
\begin{equation}\label{eq:terms}
\begin{gathered}
DD(\theta) = \frac{P_{DD}(\theta)}{N(N-1)}\\[6pt]
DR(\theta) = \frac{P_{DR}(\theta)}{N N_{R}}\\[6pt]
RR(\theta) = \frac{P_{RR}(\theta)}{N_{R}(N_{R}-1)}, \\[6pt]
\end{gathered}
\end{equation}
where $N$ and $N_R$ are the total number of data and random points in the survey volume, respectively and $P_{DD}(\theta)$, $P_{DR}(\theta)$, and $P_{RR}(\theta)$ represents the total pair counts in each separation $\theta \pm \delta \theta$ bin for the data-data, data-random, and random-random pairs, respectively.  The size of the bin is determined by the sample size; we calculate our correlation function with eight spatial bins, spaced logarithmically between 0.16'' (4 pixels) and 200'' (5000 pixels), corresponding to spatial scales of 4.8 pc to 9.6 kpc, the largest scale we can investigate with our data.  Bin numbers were selected as a compromise between resolution and total number of clusters available to be sampled within each bin.  When computed as above, a random distribution of an unclustered population will result in a flat correlation $\omega(\theta)=0$, while a clustered distribution will have $\omega(\theta)>0$.  For hierarchical structuring, a general trend of decreasing $1+\omega(\theta)$ with radius is expected \citep{gomez93,larson95}.  This correlation function has also been successfully used for characterizing the clustering behavior of both stars and stellar clusters \citep[e.g., ][]{gomez93,zhang01,scheepmaker09,gouliermis15a,gouliermis15b} in several nearby galaxies.  Although the formulation in Eq \ref{eq:LS} is optimized for taking into account edge effects, we still attempt to reproduce as closely as possible the geometry of the galaxy region sampled.  As NGC 628 is nearly face-on, we do not need to take into account deprojecting the data before computing the correlation function.  

It is also important that the random catalog must be large enough to not introduce Poisson error in the estimator.  This is checked by ensuring that the $RR$ pair counts in the smallest bin are high enough such that Poisson errors are negligible in the total error budget.  As a result, the number of random points need to be greater than the data points.  We achieve this for a random catalog that is 100 times greater than the size of the cluster sample.  The uncertainties in the measurements of $\omega(\theta)$ in each radius bin are estimated as a Poisson error.  

\subsection{Application of the Clustering Estimator}
\begin{figure}
\epsscale{1.2}
\plotone{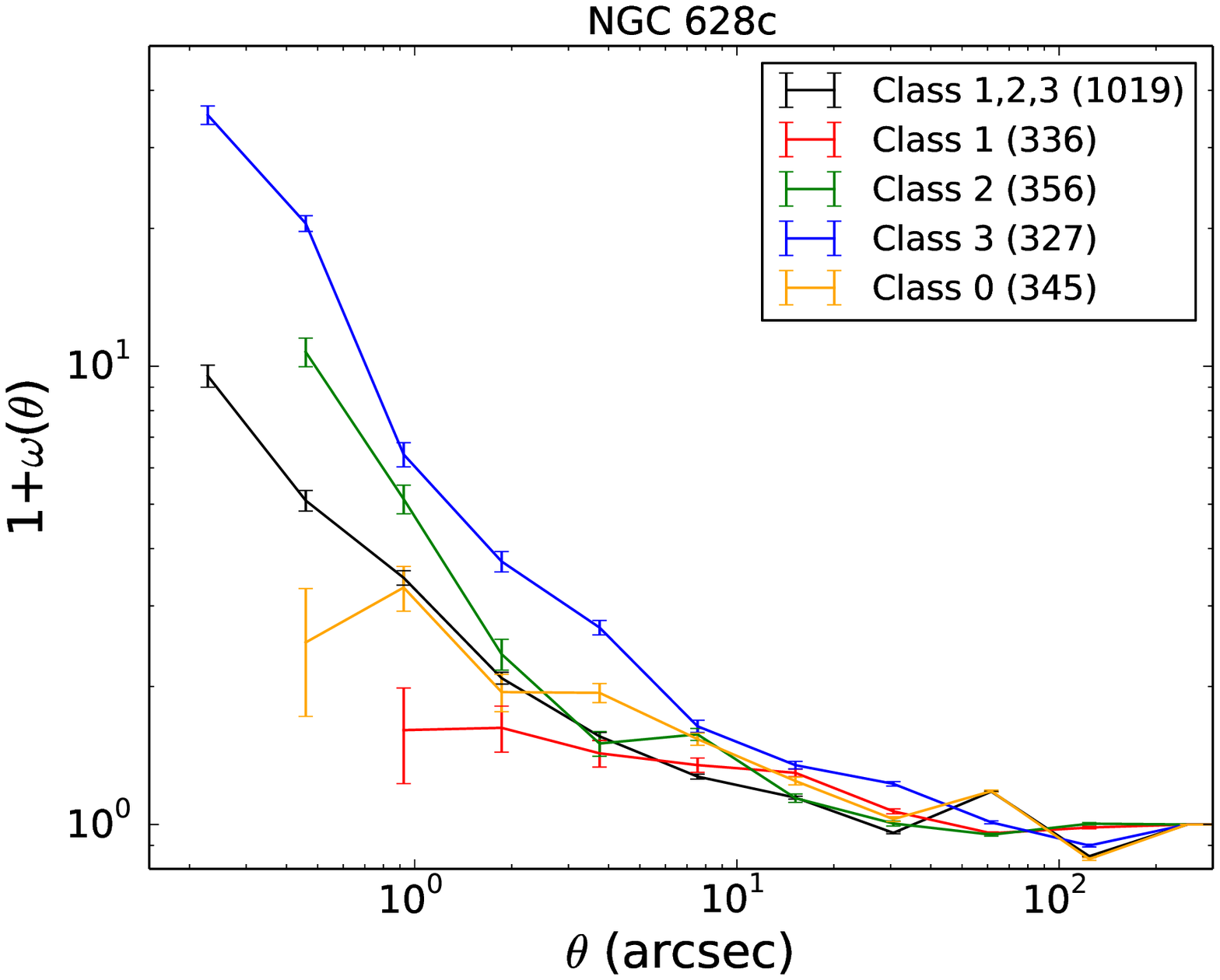}
\plotone{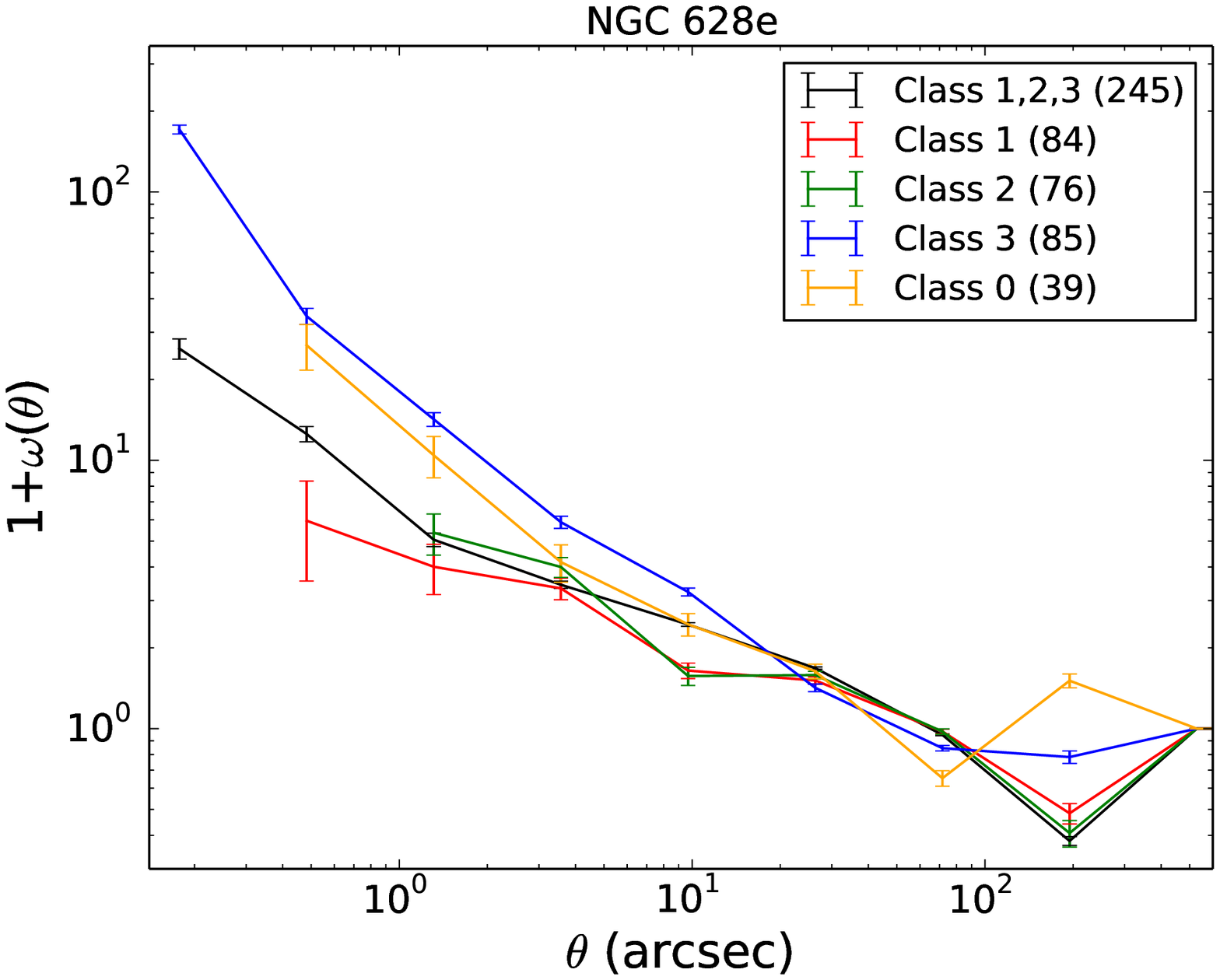}
\plotone{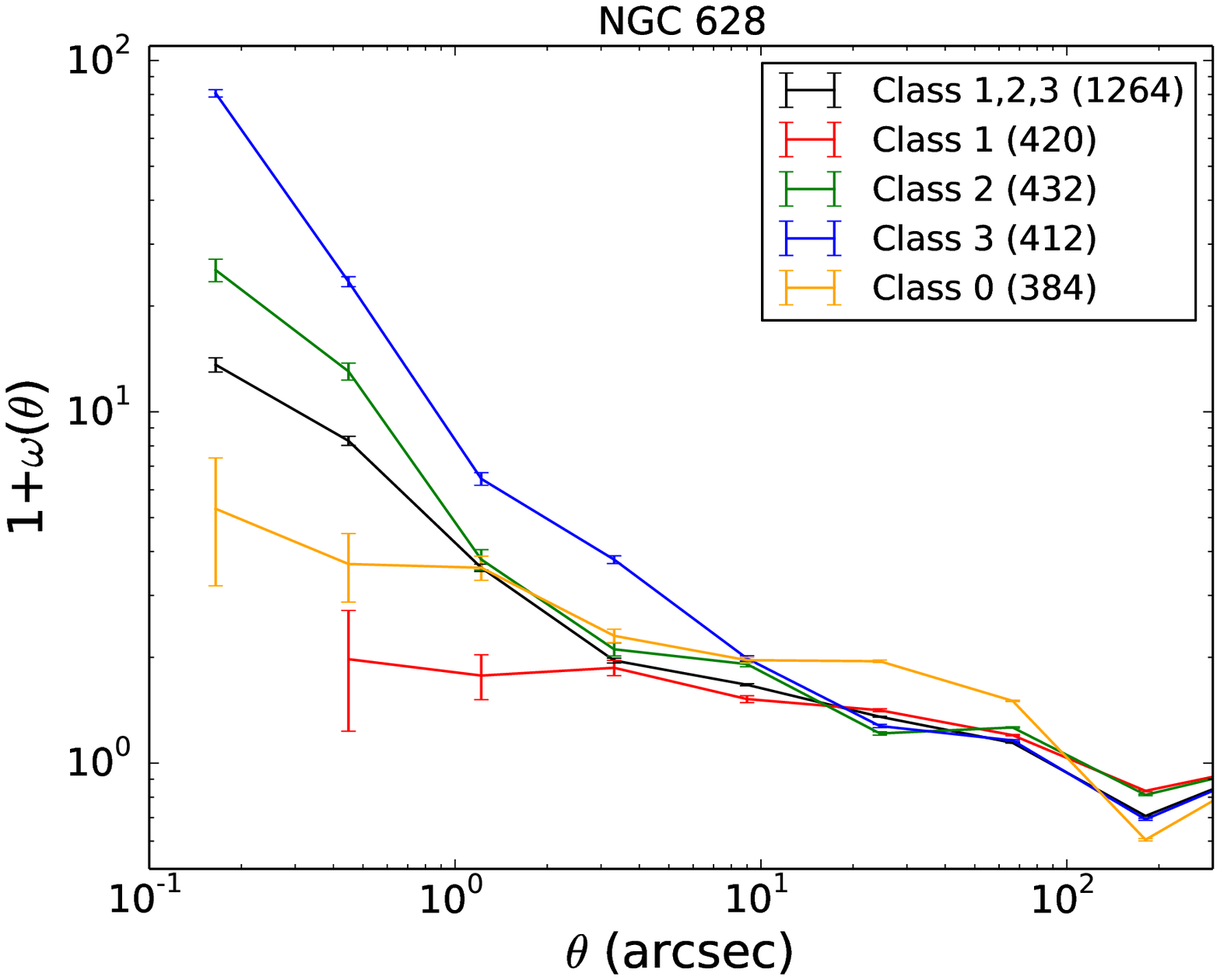}
\caption{
The two-point correlation function $1+\omega(\theta)$ for the clusters of the central pointing, NGC~628c (top), east pointing, NGC~628e (middle), and mosaic, NGC~628 (bottom) as a function of angular distance (arcsec).  The colors represent the classification of each cluster, as defined in Section \ref{sec:clusterselection}.  The numbers in parentheses show the number of clusters in each classification.  
\label{fig:2pcf}}
\end{figure}

Applying the two-point correlation function to our star clusters helps us to identify common age structures and derive the correlation length as a function of age and location.  Figure \ref{fig:2pcf} shows the correlation function for both pointings of NGC 628 as well as the mosaic.  It is interesting to note that the Class 1 (centrally concentrated) clusters have a nearly flat relationship across all spatial distributions, while the strongest clustering is seen in Class 3 (multiple peak) sources.  The correlation for Class 0, which potentially may include also spurious sources, is flatter than that for our most structured cluster but still has a distribution that is more fractal (i.e., steeper exponent) than class 1.  We will discuss in Section \ref{sec:powerlawfit} and \ref{sec:agecuts} the differences between the correlations for the separate classes of clusters, and how we will attempt to single out the Class 0 clusters without visual inspection.  

As mentioned above, the LS estimator is minimally sensitive to the random sample size and is effective at handling both edge and masking effects, as found in \citet{landy93} and \cite{kerscher00}.  Using the central pointing, NGC~628c, to test how changing the geometry and masks within the random data sample is going to affect the results, we found that a random catalog of the same observing geometry as the real data with a random sample size of 10,000 objects already converges to the same correlation function values as a random catalog with 500,000 sources.  As a sanity check, we create random fields with 1200 and 70,000 objects and compute the correlation function between the two catalogs and find that the computed $\omega(\theta)$ is zero across all angular scales.  

\begin{figure}
\epsscale{1.2}
\plotone{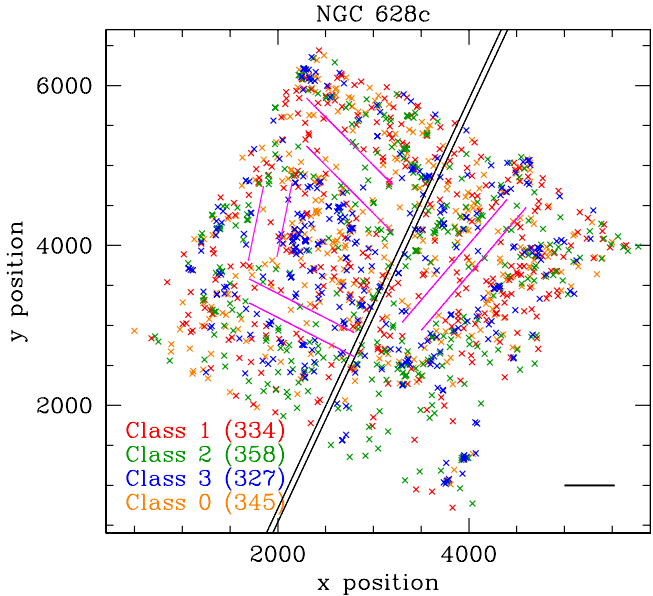}
\caption{
The positions of each cluster within NGC 628c, color coded by the classification of the clusters:  Class 1 objects are shown in red, Class 2 are shown in green, Class 3 are shown in blue, and Class 0 are shown in orange.  The numbers in parentheses show the number of clusters in each classification.  The solid black line in the bottom right represents the spatial scale of 1 kpc at the distance of the galaxy.  The two solid black lines denote the ACS chip gap and the magenta lines show the regions where we removed objects from the random catalog in order to simulate masking of dust lanes. 
\label{fig:mask}}
\end{figure}

\begin{figure}
\epsscale{1.2}
\plotone{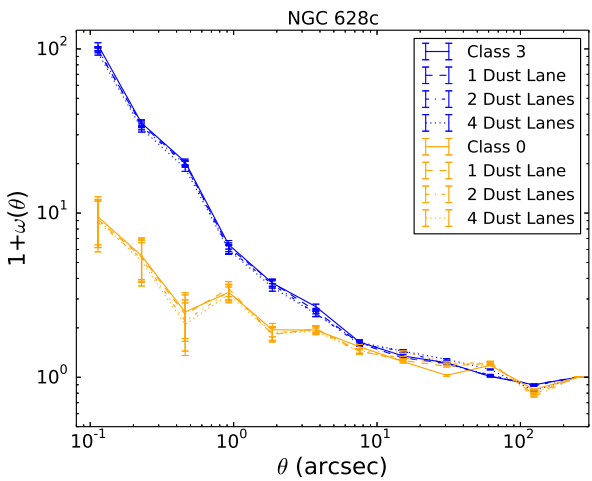}
\caption{
The two-point correlation function $1+\omega(\theta)$ for the clusters of the central pointing, NGC~628c as a function of angular distance (arcsec) for only Class 3 objects (blue) and Class 0 objects (orange) when data are removed from the random catalog representing different numbers of dust lanes.  Removing an increasing amount of data does negligibly decrease the value of $\omega(\theta)$ at small radii.  
\label{fig:2pcfmask}}
\end{figure}

We also verify that the LS correlation function is only minimally sensitive to the presence of masks within the data.  After populating the random catalog for NGC 628c, we first remove the small chip gap (see Figure \ref{fig:mask}) of several hundred pixels that runs across the length of the detector from the random catalog.  In order to test how missing data alters the correlation function, we simulated dust lanes in the random catalog that coincide with the spiral arms in the galaxy where there is a dearth of observed star clusters.  Adding anywhere between a single dust lane (about 6\% of the total area) up to four dust lanes (13\% of the total data area) have a negligible effect on the resulting correlation function, shown in Figure \ref{fig:2pcfmask}.  However, the clusters are already undersampled in the interarm regions and their removal along with the random objects in the same region, as expected, does not affect the results.  Removing the random sources from these undersampled regions results in an expected decrease of the observed clustering at small scales as it removes the data-random crosscorrelation count pairs for the clusters that reside next to these regions but does not have other major effects.

\begin{figure}
\epsscale{1.2}
\plotone{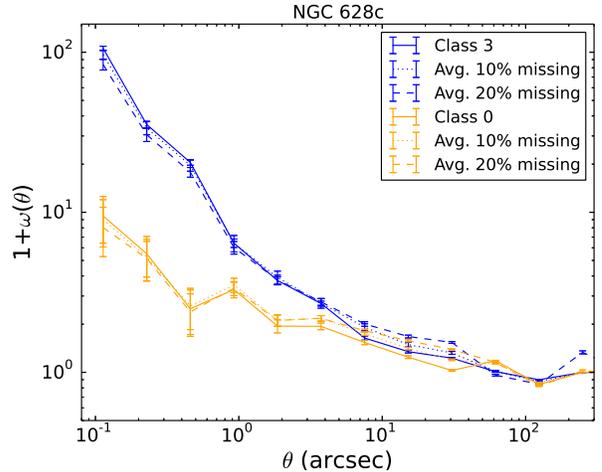}
\caption{
The two-point correlation function $1+\omega(\theta)$ for the clusters of central pointing NGC~628c as a function of angular distance (arcsec) for only Class 3 objects (blue) and Class 0 objects (orange), where we mask regions from the random catalog that corresponds to either 10 or 20\% of the total area covered by the data to see how that affects the clustering results.  Shown is the correlation result from averaging the output from 10 realizations, each with random regions excluded from the galaxy.   
\label{fig:2pcfmissingdata}}
\end{figure}

To test the effect of masks in regions that are not undersampled, we take a random catalog with the same geometry as covered by the real data and remove regions from the random catalog that cover 10 or 20\% of the total area within the galaxy and recompute the correlation function.  We keep all the data and just remove the region from the random field.  We perform this test 10 times, moving the region of missing data around the galaxy, and take the average resulting correlation function, shown in Figure \ref{fig:2pcfmissingdata}.  For areas that are outside of the dust lanes (i.e., we are not selecting regions where the clusters are particularly undersampled), we observe a slight decrease in the clustering at small spatial scales and a transition toward increased clustering at larger spatial scales.  As we are removing regions that are populated more than the dust lane regions, this can be best understood as an extreme example of removing data in regions that are not undersampled.  The increase of clustering strength at the largest spatial scales is similar to what is observed when we artificially increase the random catalog size.  These differences are not particularly significant from the statistical point of view.  

\begin{figure}
\epsscale{1.2}
\plotone{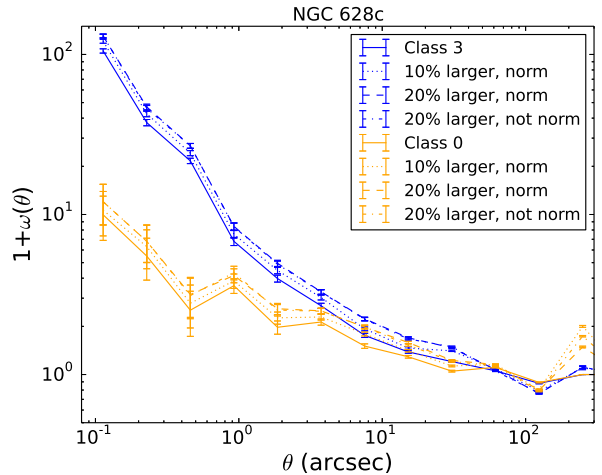}
\caption{
The two-point correlation function $1+\omega(\theta)$ for the clusters of the central pointing NGC~628c as a function of angular distance (arcsec) for only Class 3 objects (blue; the most clustered) and Class 0 objects (orange), where we compare the results for a random catalog that perfectly matches the area covered by the real data compared to catalogs that are made with a coverage area that is 10\% and 20\% larger than the real data.  In both cases, the number of sources in the random catalog is scaled accordingly to account for the increase in area coverage in order to maintain a constant surface density.  When the random catalog is larger than the area covered by the data, the correlation strength is artificially increased.  
\label{fig:2pcfgeo}}
\end{figure}

We finally test the case of a random catalog with area that is 10\% or 20\% larger than that of the actual data (Figure \ref{fig:2pcfgeo}).  For this test, we normalize the number of objects within the random catalog with the size of the random field geometry so that each field has the same surface density of random sources.  In this case, we observe an increase in the strength of the clustering (higher values for the two point correlation function).  Likewise, a random catalog that has a smaller area than the coverage of the real data will result in a decrease in clustering across all scales.  The effect of matching area and shape becomes especially important for galaxies that do not span the entire width of the chip and have to be described with ellipses for a subset of the chip coverage.  While not applicable to NGC 628, it does affect other LEGUS galaxies in the survey.

\section{Results and Analysis}\label{sec:results}

\subsection{Quantifying the Correlation Strength}\label{sec:powerlawfit}
\begin{figure*}
\epsscale{1.2}
\plotone{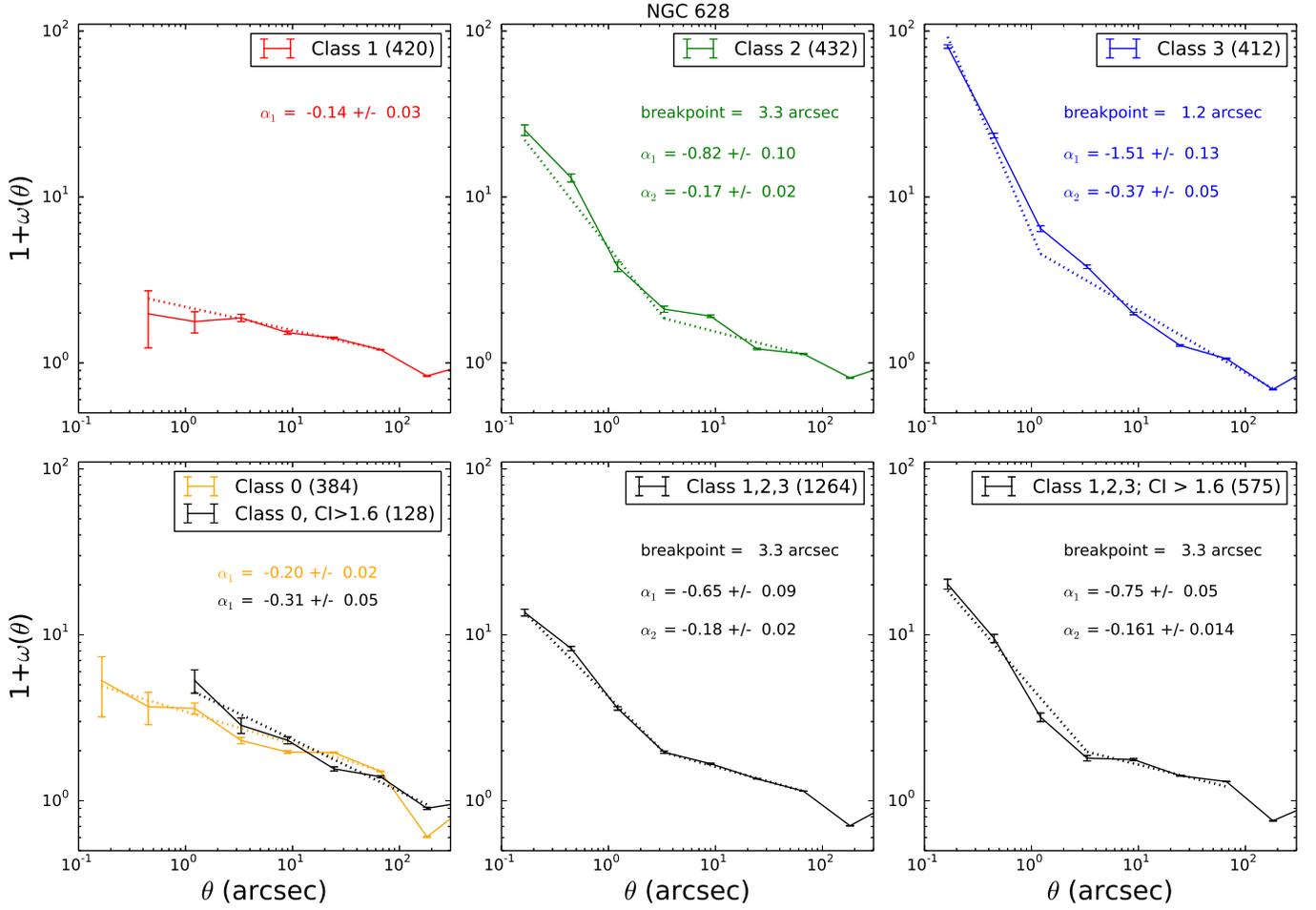}
\caption{
The two-point correlation function $1+\omega(\theta)$ for the clusters of NGC 628 as a function of angular distance (arcsec), separated into classes and fitted with broken power law, given as $\alpha_1$ and $\alpha_2$ along with uncertainties to the power law slope.  In the bottom left, we show all the Class 0 sources (yellow) and only those limited to CI $>$ 1.6 (black).  We also show the power law fit to the correlation function for Class 1, 2, and 3 clusters (bottom panel, middle) and all of the clusters with CI $>$ 1.6 (bottom right plot). 
\label{fig:2pcfpower628}}
\end{figure*}

The projected angular two-point correlation function (Eq \ref{eq:LS}) is usually well described with a power-law,
\begin{equation}\label{eq:powerlaw}
\omega(\theta) = A_{\omega}\theta^{\alpha},
\end{equation}
where the slope $\alpha$ measures the strength of the clustering and the amplitude $A$ measures the correlation length of the clustering; we use both to determine if the clustering is consistent with being scale-free.  In a fully hierarchical model, the clusters will be correlated with other clusters from other groupings on larger scales, together forming a group higher up in the hierarchy of star formation.  This would result in a smooth decline of $1 + \omega(\theta)$ in both radius and age.  Figure \ref{fig:2pcfpower628} shows the two-point correlation function after combining both pointings in NGC 628; we show each cluster class in a separate panel, and fit each correlation function with a broken power law (except for Class 0 and 1 which are better fit with a single power law).  We compute the fit and determine the slopes using the Levenberg-Marquardt non-linear least square minimization fit, where the functional form is given as, 
\begin{equation}
\log [1+\omega(\theta)] = \left\{
  \begin{array}{lr}
    A_1 + \alpha_1 \log(\theta) &  : \log(\theta) < \beta \\
    A_2 + (\alpha_1 - \alpha_2)\beta  + \alpha_2 \log(\theta) &  : \log(\theta) > \beta,
  \end{array}
\right.
\end{equation}
and where the breakpoint $\beta$ is the logarithm of the position of the separation break along the x-axis (spatial scale), $A_1$ and $A_2$ are the clustering amplitudes before and after the break, and $\alpha_1$ and $\alpha_2$ are the slopes to the power law before and after the breakpoint, respectively.  Slopes, amplitudes, and break-points are free parameters in the fit.  The fitted parameter results for NGC 628 are listed in Table \ref{tab:1}.  

For a hierarchical (scale-free or self-similar) distribution, the total number of clusters $N$ within an aperture increase with radius $r$ as $N \propto r^{\alpha+ 2}$, where $\alpha$ is related to the (two-dimensional or projected) correlation fractal dimension D2 $= \alpha + 2$ \citep[e.g., see ][]{falgarone91}.  The distribution of star formation and interstellar gas over a large range of environments is observed to show a (three-dimensional) fractal dimension of D3 $\sim$ 2.3 \citep[e.g., ][]{elmegreen96,elmegreenelmegreen01}, in agreement with the predicted fractal dimension for the density structure of a turbulent ISM \citep{federrath09}.  Projection of this three-dimensional fractal on a plane will result in a two-dimensional fractal dimension D2 = D3 -- 1 if the perimeter--area dimension of a projected 3D structure is the same as the perimeter--area dimension of a slice of the structure \citep{elmegreen04}.  The two-point correlation function results are qualitatively identical to a power spectra analysis.  However, as shown in simulations by \citet{gouliermis14}, the D2/D3 conversion is not trivial.  

As can be seen in the power law fits to NGC 628 in Figure \ref{fig:2pcfpower628}, the power law index recovered ranges from $\alpha=-0.14$ to $-1.51$, covering spatial scales from 5 pc to nearly 5 kpc.  If we take the weighted mean of the measured indices before the break point at 3\farcs3, we recover an average slope of $\alpha \sim -0.8$, in agreement with what is expected from a hierarchical distribution of gas.  After the break point, we observe a dramatic decrease in the fractal distribution, informing us that the spatial distribution of the clusters (excluding Class 1 and 0 sources) is systematically less clustered at larger spatial length-scales.  Results of a study of the hierarchical structure of star formation within a subset of 12 LEGUS galaxies by \citet{elmegreen14}, suggests that hierarchically structured star-forming regions are common unit structures, which can be several hundred parsecs, but also that observed self-similarity of young stellar structures down to parsec scales is indicative that individual star clusters form stellar groupings on larger scales.  

Given the 2\farcm7 FOV of WFC3/UVIS, the largest physical size that can be probed for a single pointing in NGC 628, is 7.8~kpc.  A pixel scale corresponds to a physical size of 1.90 pc at the distance of 9.9~Mpc, with a typical star cluster around 5--10 pc in diameter.  Each of the classes is best-fit with a broken power law at 3\farcs3, corresponding to a spatial scale length of 158~pc at a distance of 9.9~Mpc, except for Class 1 clusters, best-fit with a single power law.  The break in the power law corresponds to the observed line-of-sight thickness of the galactic disk (see Section \ref{sec:appbreak} and \ref{sec:discussion} for discussion).  Class 3 clusters exhibit the break below the rest of the clusters at 1\farcs2 in addition to exhibiting a much steeper slope across all scale lengths.  The exact location of the breakpoint naturally depends on the number of bins used to calculate the correlation function (See Section \ref{sec:clusterselection}).  Past the 3\farcs3 break, all the classes show a correlation function with a power law index of $\alpha \sim -0.2$, indicating that any occurring clustering has considerably weakened by this length scale, lasting until a few kpc.  The clustering observed increases below this breakpoint, albeit to varying degrees between the cluster classifications.  These results indicate that star formation is clumpy and the stellar clusters form highly clustered distributions for spatial separations of $r$ less than the break point.  For spatial scales above the breakpoint, the clustering observed becomes more homogeneous and less clustered.  

\begin{deluxetable*}{lcccccc}
\tabletypesize{\scriptsize}
\tablecaption{Power-Law Parameters of NGC 628 \label{tab:1}} 
\tablecolumns{7}
\tablewidth{0pt}
\tablehead{
\colhead{Class}	& 
\colhead{Number}	& 
\colhead{$A_1$}				&  
\colhead{$\alpha_{1}$} 	&
\colhead{$\beta$}   	&
\colhead{$A_2$}				&  
\colhead{$\alpha_2$} 	
\\
\colhead{}	& 
\colhead{}	& 
\colhead{}	&  
\colhead{} 	&
\colhead{(arcsec)}   &
\colhead{} 	&
\colhead{}   
}
\startdata 
Class 1 					& 420 		& 2.24(3)	& $-0.14(3)$ 	& \nodata&\nodata& \nodata \\
Class 2 					& 432 		& 5.31(3)	& $-0.82(10)$ & 3.3 		& 2.76(2)	& $-0.17(2)$ \\
Class 3 					& 412 		& 8.17(7)	& $-1.51(13)$ & 1.2 		& 3.4(2)	& $-0.37(5)$ \\
Class 0 					& 384 		& 3.44(5)	& $-0.20(2)$ 	& \nodata&\nodata& \nodata \\
Class 0; CI$>1.6$ 	& 128 		& 	4.8(1.2)& $-0.31(5)$ 	& \nodata&\nodata& \nodata \\
Class 1, 2, 3 			& 1264 	& 3.91(19)& $-0.65(9)$ 	& 3.3 		& 2.29(2)	& $-0.18(2)$ \\
Class 1, 2, 3; CI$>1.6$ & 575 & 3.9(3)	& $-0.75(5)$ 	& 3.3 		& 2.26(2)	& $-0.161(14)$ 
\enddata
\tablecomments{
Columns list the 
(1) Classification of stellar clusters -- Class 1: symmetrical clusters; Class 2: asymmetrical clusters; Class 3: multiple peak clusters; and Class 0:  non-visually identified clusters, 
(2) Number in each classification, 
(3) Amplitude $A_1$ of the angular correlation function before the breakpoint, 
(4) Slope $\alpha_1$ of the angular correlation function after the breakpoint, 
(5) Location of the break point $\beta$.  Cluster classifications that are best-fit with a single power law do not have a breakpoint, 
(6) Amplitude $A_2$ of the angular correlation function past the breakpoint, and
(7) Slope $\alpha_2$ of the angular correlation function after the breakpoint.  Numbers in parentheses indicate uncertainties in the final digit(s) of listed quantities, when available. 
}
\end{deluxetable*}

\begin{figure}
\epsscale{1.2}
\plotone{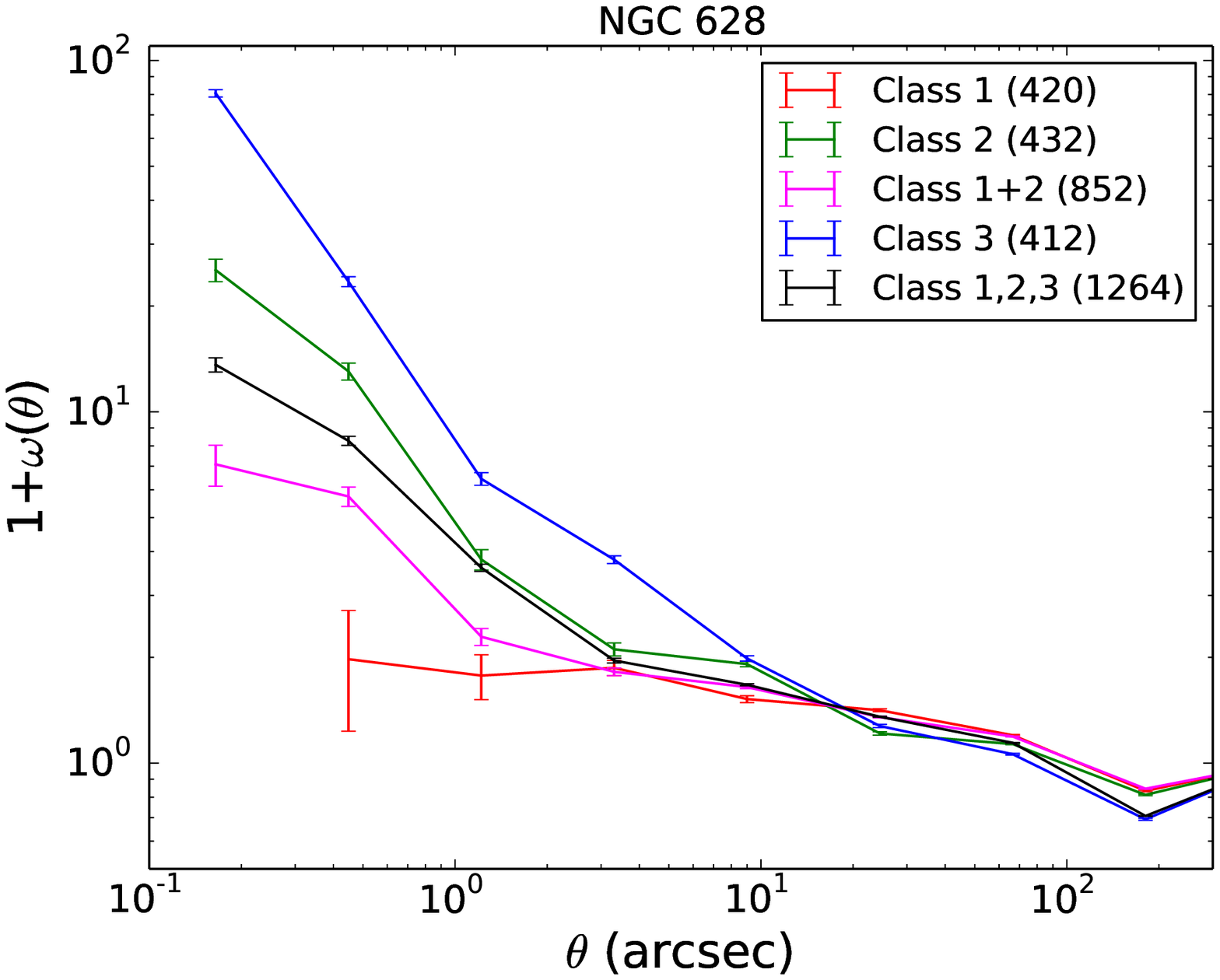}
\caption{
The two-point correlation function $1+\omega(\theta)$ as a function of angular distance (arcsec) for the cluster classifications where we compare what we believe to be bound clusters (Class 1 and 2; magenta) to the behavior observed for the combination of all the Class 1, 2, and 3 (black) clusters.  
\label{fig:2pcf_class12}}
\end{figure}

In order to address how clusters behave versus associations, we compare Class 1+2 (bound) clusters to just Class 3 clusters in Figure \ref{fig:2pcf_class12}.  We believe that Class 1 and 2 are gravitationally bound clusters in different states of relaxation whereas Class 3 (multiple peak) clusters quite possibly be unbound associations.  We see that the clustering behavior of the Class 1+2 is different than what is observed for just the Class 3 clusters alone.  These results indicate that bound clusters (class 1 and 2) have a correlation behavior that is weaker, corresponding to a different spatial distribution when compared to what we observe for Class 3 and the compilation of Class 1, 2, and 3 clusters combined.  This strengthens our results that as (bound) clusters relax, they become less clustered and behave differently than what is observed for associations.  

\subsubsection{Break in the Power Law}\label{sec:break}
Excluding Class 0 and 1 sources, distributions are described with a broken power law with the break point occurring at 158~pc averaged between all the genuine clusters (class 1, 2, and 3), though Class 3 clusters lose their self-correlation at much shorter scale lengths -- 58 pc.  In fact, we may expect a smaller characteristic scale for the younger clusters and a larger one for older clusters, if clusters and stars represent the same hierarchy \citep{efremov95}.  Class 1 clusters are the least self-correlated of all the cluster morphologies with a nearly flat exponent, implying that the distribution of these clusters is nearly uniform across all scales.  The flattening of the exponent found after the break point for classes 2 and 3 is comparable to the uniform distribution of Class 1, implying the loss of self-correlation distribution for these types of clusters after the break point at the larger scale.  For the cumulative correlation function of class 1, 2, and 3, we still see strong self-correlation with a break point below 158 pc, suggesting that the Class 2 and 3 clusters are dominating the small-scale correlation.  


A similar break in power spectra within other galaxies has been observed and interpreted as due to the line of sight thickness of galactic disks \citep{lazarian00,elmegreen01,odekon08,block10} when the line-of-sight depth is smaller than the transverse size.  A study of the brightest stars in the LMC, M31, and M33 by \citet{odekon08}  identified a transition in the correlation function to a higher correlation dimension (weaker clustering), with the transition marking the large-scale regime where disk geometry and dynamics set the scale for structure.  This transition from the small-scale three-dimensional turbulence to large-scale two-dimensional disturbances in the disk, observable with a transition to a more shallow power law at the transition, has been detected for the HI gas in the LMC with a transition occurring at 180 pc by \citet{padoan01,elmegreen01} and 290 pc by \citet{kim07}.  We may very well be seeing the analogous transition within our young stellar populations.  


\subsubsection{Dependency of Results on Bin Numbers}\label{sec:appbreak}
\begin{figure}
\epsscale{1.2}
\plotone{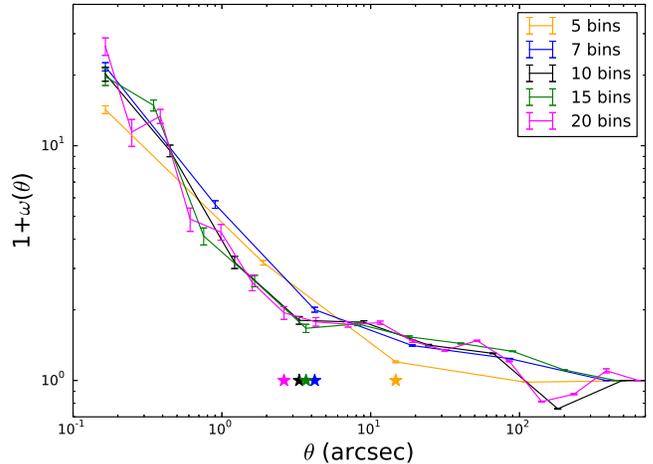}
\caption{
The correlation function computed using different widths of annuli, ranging from 5 to 20 bins across the spatial scale we are investigating.  The star symbols show the location of the breakpoint for each result.  
\label{fig:binnum}}
\end{figure}

The choice of annuli width used to calculate the correlation function in Eq. \ref{eq:LS} will naturally influence the results and the exact location of the breakpoint of the correlation behavior.  In order to ensure adequate sampling of clusters in each bin across the scale length that we are investigating, our bin numbers were selected such that the smallest angular bin contained a minimum of two clusters, as the smallest separation size is where our data are the sparsest.  Figure \ref{fig:binnum} shows how changing the bin size from five bins to 20 bins moves the location of the breakpoint, averaged over Class 1, 2, and 3 clusters, between 2 to 14 arseconds (96 pc to 672 pc, respectively).  Decreasing the bin numbers aids to smooth out occurring variations at both small and large angular separation.  We can see that the break point only starts to disappear at the coarsest sampling of 5 bins, highlighting that galaxies with a large number of clusters available are needed to provide statistically sound results.  In order to see the breakpoint at the disk thickness, we need to be able to resolve the equivalent scale.  

\subsection{Age and Concentration Index}\label{sec:agecuts}
\begin{figure*}
\epsscale{1.2}
\plotone{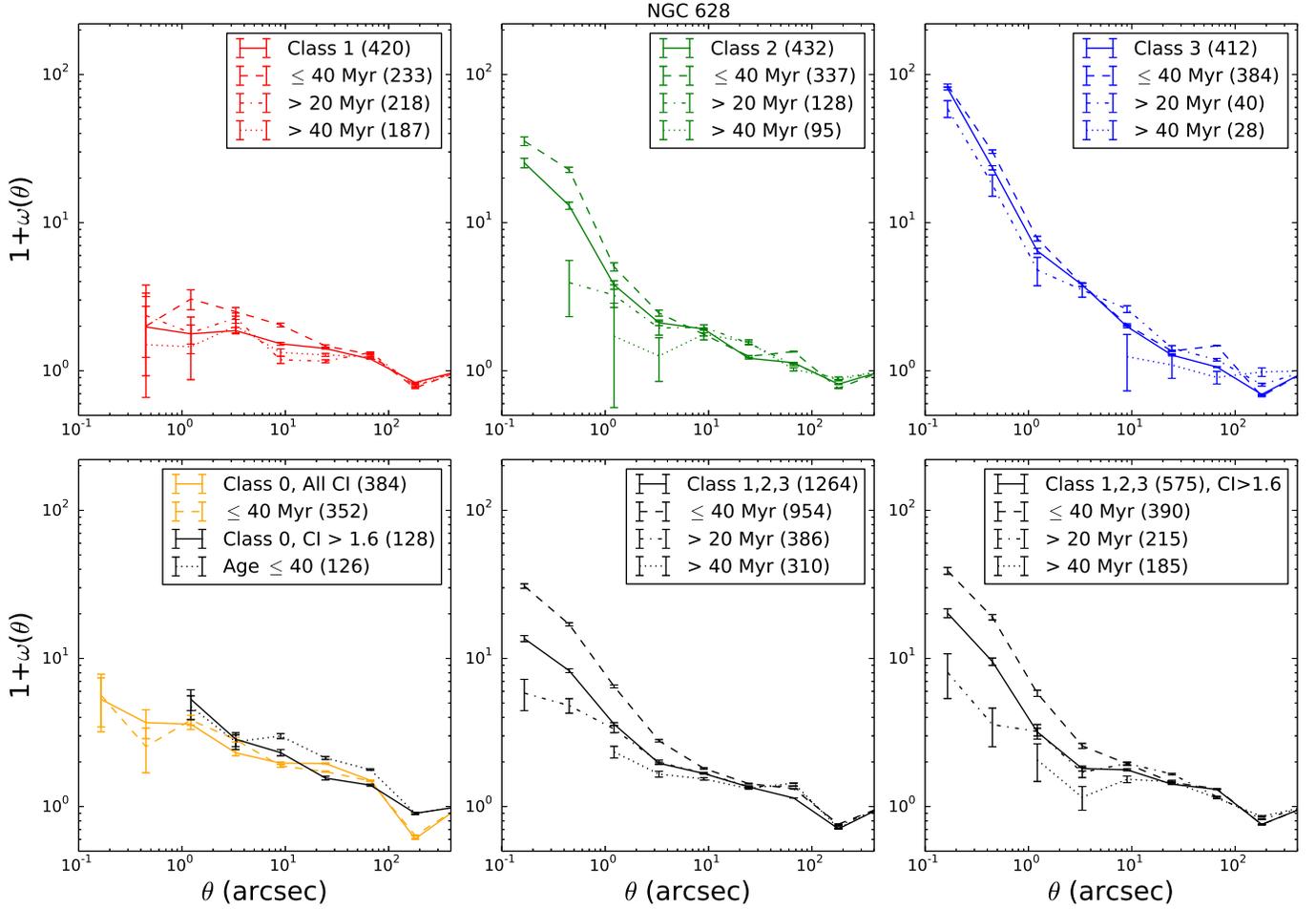}
\caption{
The two-point correlation function $1+\omega(\theta)$ as a function of angular distance (arcsec) for the cluster classifications in NGC~628 as shown in Figure \ref{fig:2pcf} along with the subset of clusters with ages below/above 40~Myr and those above an age of 20 Myr.  The strength of clustering increases when we only consider the youngest clusters within each classification.  For Class 0 sources, we also show how the CI (compact vs broad) values influence the clustering, where the broad Class 0 sources (CI $>$ 1.6 mag) appear to have a slight increase in the clustering strength.  Diving the genuine clusters (i.e., Class 1, 2, and 3) by CI value does not change the amount of clustering.  
\label{fig:2pcfage}}
\end{figure*}

The correlation function of Class 0 objects tends to be very shallow, confirming that Class 0 objects contain a compilation of every possible object available.  In order to attempt to identify only the star clusters within Class 0 objects, we take very conservative cuts in CI and then further divide into age bins.  The youngest objects that are genuine star clusters should exhibit an increase the strength of the power-law in the correlation function.  We see this in the correlation function for Class 3 clusters, which on average are younger and more clustered compared to Class 1 clusters.  In order to test if the correlation does change with cluster age, we divide our clusters by age, starting at 10 Myr, and recompute the correlation function.  Figure \ref{fig:2pcfage} shows the correlation function for all cluster classifications as well as only the clusters within each classification that are younger than 40~Myr, which is the age where we see the most noticeable transition from clustered to non-clustered.  Excluding Class 0 and 1 clusters, the correlation function does increase when we only consider the youngest clusters in each class.  Separating the broadest sources (CI $>$ 1.6 mag) for the combination of the Class 1, 2, and 3 sources from their narrower counterparts (far right, bottom panel of Figure \ref{fig:2pcfage}) only marginally increases the clustering strength.  The shallowness of the slope for the Class 1 sources is not entirely surprising as these are the oldest and most massive clusters in the galaxy, where we possibly are viewing a subset of clusters that have dispersed and have lost their clustered, nonhomogeneous distribution.  Class 3 sources, which are on average the youngest in the galaxy, are systematically the most clustered class observed.  

As can be seen in our analysis, at 20 My we see a transition at which the clustering structure disappears and the clusters display a more homogeneous, non-clustered distribution, in place by ages of 40 Myr.  However, the clustering strength of Class 3 clusters are only minimally effected with an increase in age.  All star clusters regardless of their classification tend to be randomly distributed above this age, where random motions and shear effects can explain the observed data trends.  The flatness of the correlation function of the age distribution for Class 1 clusters as compared to Class 3 clusters can be evidence of cluster dissolution as we see a dramatic decrease in the total strength of the clustering with increasing age. 



We also show the Class 0 sources in Figure \ref{fig:2pcfage} divided into concentration index bins above and below CI of 1.6 mag, where the broader sources (CI $>$ 1.6 mag) do show evidence of increased clustering.  We do not show the Class 0 sources for ages older than 40 Myr as there is a dearth of available data.  When we combine the youngest and broadest sources, limiting our selection to our very conservative cut of CI $\geq1.6$ mag and divided into extremely young ($\leq10$~Myr) and young ($\leq40$~Myr) age bins, we do not see a difference in the strength of the clustering.  However, out of all the CI $\geq1.6$ mag sources (128), 98\% have ages of $\leq40$~Myr, which indicates that sources with high values of CIs are sources with younger ages.  

\begin{figure}
\epsscale{1.2}
\plotone{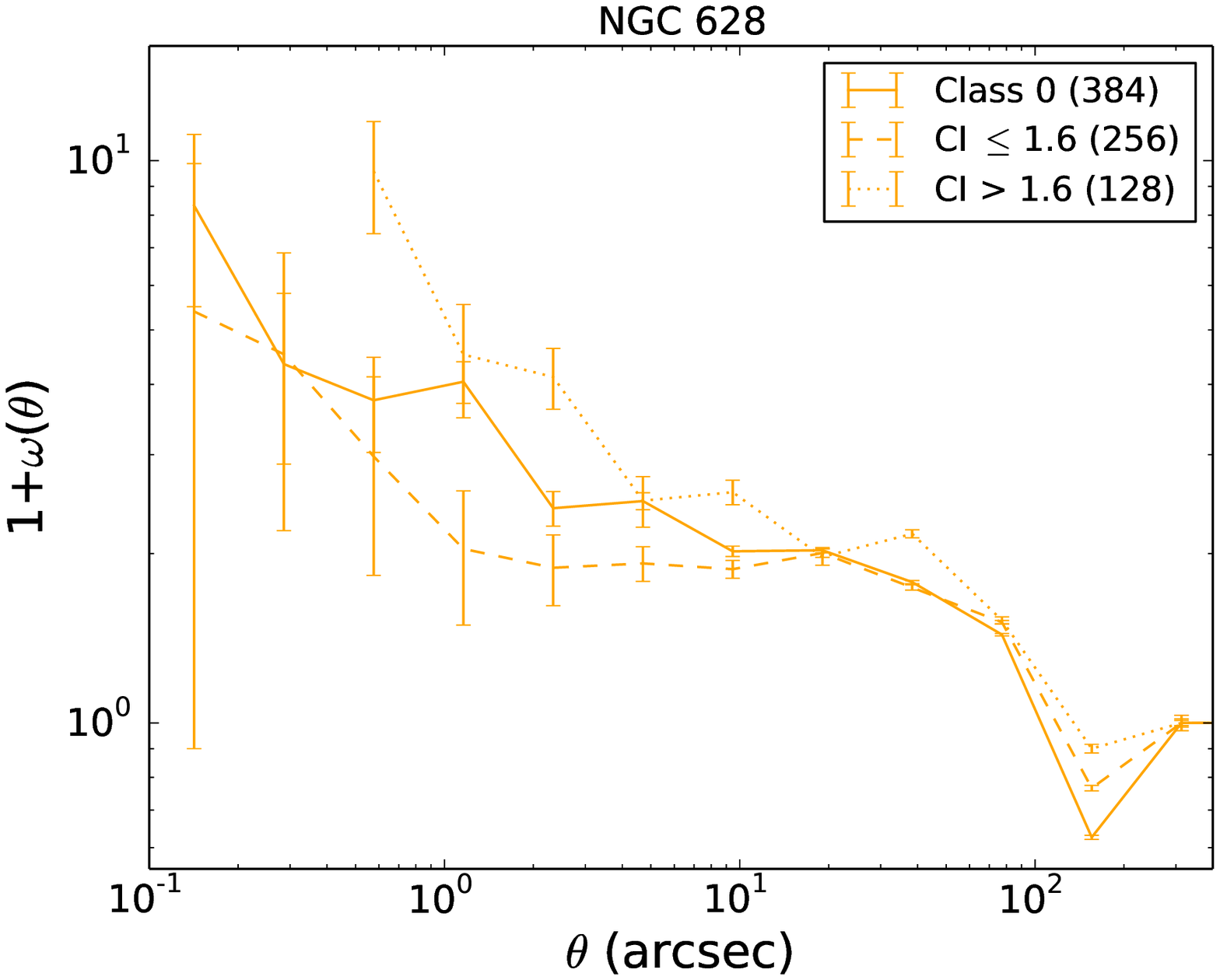}
\plotone{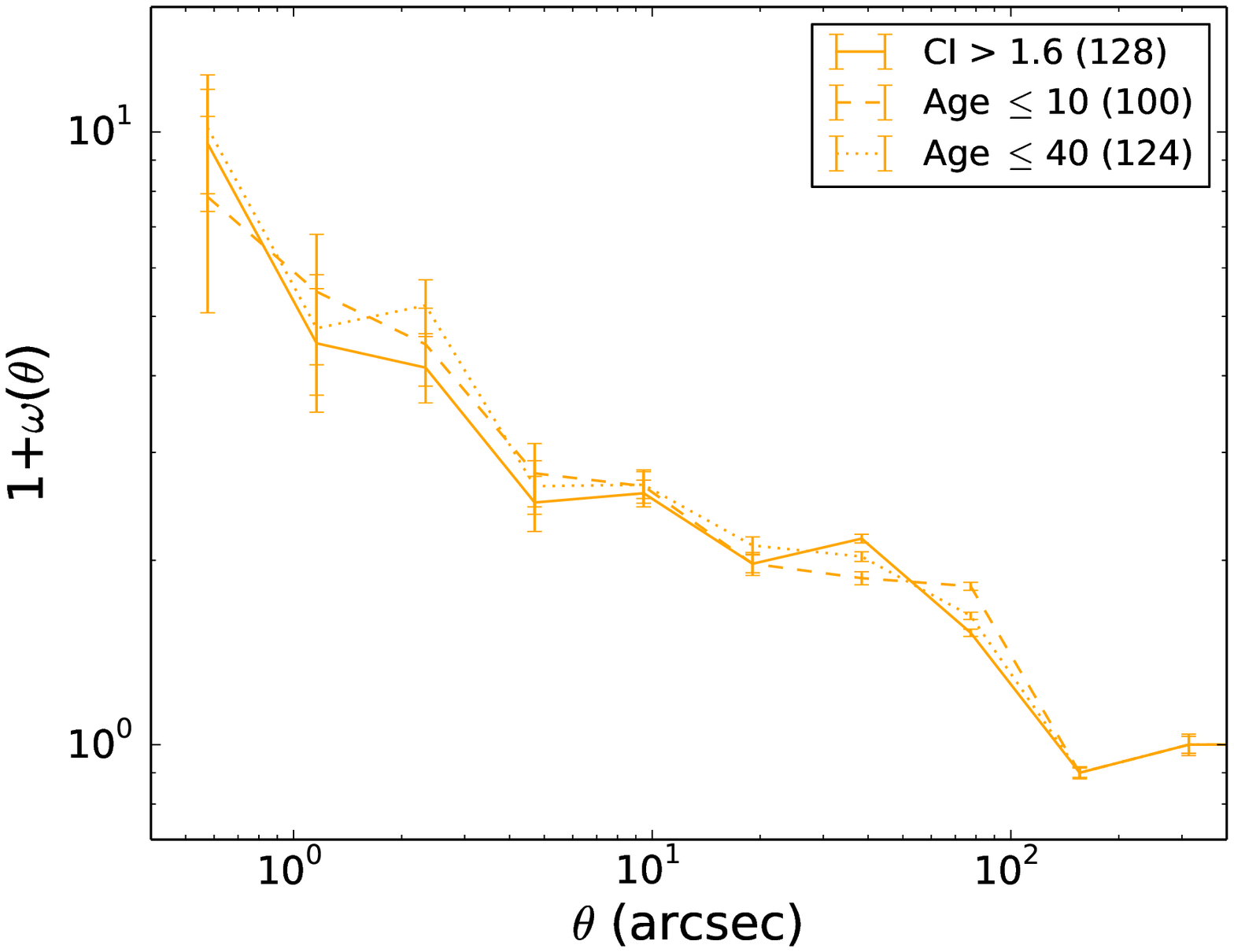}
\caption{
Top:  The correlation function for Class 0 sources, divided into bins of high and low-CI.  We classify compact clusters as those with a CI mag less than 1.6 and broad clusters as those with a CI$>1.6$.  The compact clusters show less evidence of clustering with increasing spatial scale for where we have data, which is to be expected if the broadest objects are clusters.  
Bottom:  The two-point correlation function $1+\omega(\theta)$ for Class 0 clusters that are above a CI cutoff of 1.6, divided into bins of ages below 10 and 40~Myr.  While genuine clusters should be identified as the younger clusters of the broader CI objects, showing more evidence of clustering at smaller spatial scales, the lack of older clusters in the CI $>$ 1.6 bin makes the correlation function nearly identical between all the ages.  
\label{fig:2pcfageCI}}
\end{figure}

As we saw in the power law fits of Section \ref{sec:powerlawfit}, Class 0 sources are described with a fairly shallow power law of index $\alpha = -0.20$, which is not as steep as expected if the objects are part of a larger hierarchical distribution.  However, the clustering does increase when we limit our selection to class 0 clusters with CI $>$ 1.6, as can be seen in Figure \ref{fig:2pcfage}.  If we repeat the power law fits to the CI $> 1.6$ clusters, we find the power law index increases to $\alpha = -0.31$, up from $\alpha = -0.20$.  This is still not as steep as the slope observed for Class 2 or 3 clusters, however, it compares to slopes measured by other star cluster studies in the range of $-0.4\ \rm{to} -0.8$ \citep{scheepmaker09} and $-0.7\ \rm{to} -1$ \citep{zhang01}.  

The age where we start to see a decrease in the clustering of stellar clusters, 40 Myr, is significantly younger than the age at which the clustering structure of stars starts to dissipate \citep[300 Myr, for eg.,][]{bastian11b}.  The difference in age can be explained if clusters lose their substructure through a combination of both dissolution and the migration.  While stars maintain their individuality over their lifetime, clusters can dissolve or evaporate over a much shorter timescale, implying that a cluster may lose its identity while the stars that compose it do not.


\subsection{Mass}
We investigate the role that the mass of stellar clusters play in the hierarchy, formation, and dispersion of stellar structures.  

\begin{figure*}
\epsscale{1.2}
\plotone{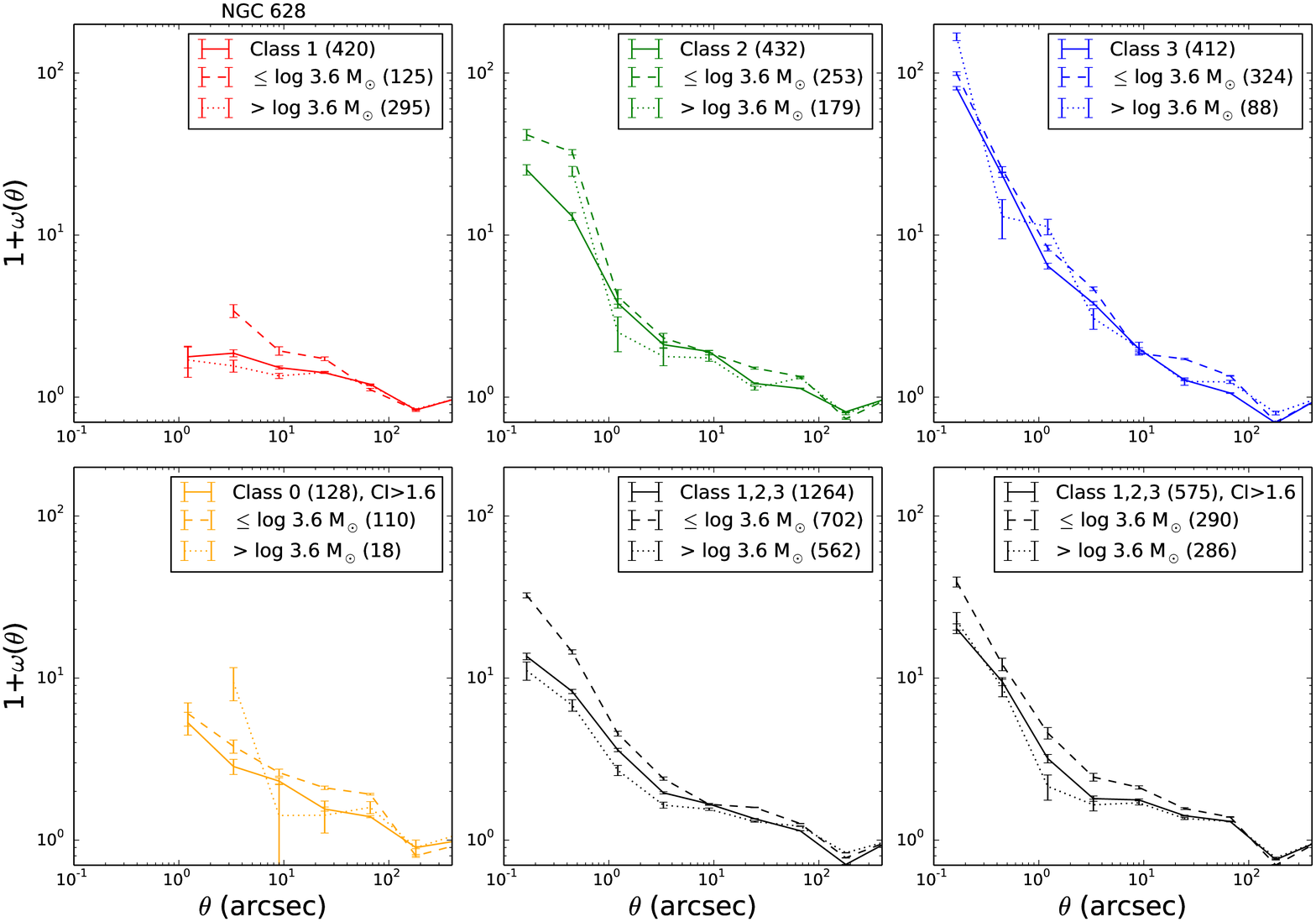}
\caption{
The two-point correlation function $1+\omega(\theta)$ as a function of angular distance (arcsec) for all the cluster classifications in NGC~628 as shown in Figure \ref{fig:2pcf}, divided into bins below (dashed line) and above (dotted line) the mass big of $\log(M/M_{\odot}) = 3.6$.  For the Class 0 sources, the sources with CI $>$ 1.6 mag are shown in orange, where we see a large increase in the clustering for the massive sources.  
\label{fig:2pcfmass}}
\end{figure*}

Figure \ref{fig:2pcfmass} shows how the clustering changes when we divide the clusters in the sample into bins of high and low mass, where we choose the value of $\log(M/M_{\odot}) = 3.6$ to have adequate numbers of clusters sampling each bin across all cluster classifications.  For Classes 1 and 2, we notice slightly stronger correlations for the lower mass clusters, where the correlation strength starts to increase for the lower mass clusters at the smallest spatial scales.  The strength of the clustering for different masses is not significantly different for Class 3 clusters.  Combining all the genuine clusters (bottom middle and right plots in Figure\ref{fig:2pcfmass}) accentuates the increased clustering strength for lower mass clusters.   We do recognize that clusters with $\log < 3.6\ M_{\odot}$ may be biased by stochastic IMF sampling, resulting in unreliable mass measurements at the low-mass end.  This makes it difficult to draw specific conclusions on how the mass affects clustering and is further compounded by the fact that we lack clusters populating the lowest spatial scales on the correlation function.  For the Class 0 sources that are limited to CI $>$ 1.6, a large increase in clustering is observed for the large mass bin of $\log(M/M_{\odot}) > 3.6$.  If we consider all the class 0 sources regardless of CI value, this same jump is not seen and there is no significant difference between the low and high-mass sources.  This increase in clustering for the more massive clusters is not seen for any of the other cluster types within NGC 628.  

\section{Discussion}\label{sec:discussion}
The hierarchical morphology in stellar ensembles has been investigated by \citet{elmegreen14} with 12 LEGUS galaxies, including NGC 628.  The power laws of star forming regions suggest that hierarchically structured regions are represented with a common unit length of a few hundred parsecs.  The observed power law of hierarchical structure of star forming complexes observed in galaxies is consistent with the model that star formation is regulated by turbulence where gas compression form successively smaller clouds within larger clouds \citep[see ][]{vazquezsemandeni09,elmegreen10,kritsuk13,elmegreen14}.  The hierarchical nature of star formation will likewise drive the observed hierarchy of young stars, and therefore, the distributions of young stellar clusters.  The same processes will likely form a secondary correlation for age, making larger regions older in proportion to the turbulent crossing time \citep{efremov98}.  These results indicate that the young stellar clusters that we observe should also be imprinted with the hierarchical structure of the natal cloud structure from which they are born.  The influence of the self-similar nature of star formation is demonstrated on the distributions of resolved massive young stars on galactic scales and the power laws in their parameters correlations, as shown for NGC 6822 \citep{gouliermis10} as well as with the much deeper LEGUS data for NGC 6503 \citep{gouliermis15b}.    Our results are consistent with the study of the luminosity function of the young stellar populations in NGC 628 by \citet{adamo15b}, found to be well-described by a power law distribution with index close to $-2$, suggesting that clusters form in a turbulence-driven fractal ISM. 

Excluding Class 0 and 1 sources, our correlation functions for NGC 628 are best described with two power laws connected at a break point corresponding to 158~pc, though this break is much shorter for Class 3 clusters.  The thickness of the galactic disk of NGC~628 is around 0.25~kpc \citep{peng88,ma98}, a factor of two different from our measured break point in the power spectrum.  Previous results of power spectra in galactic observations have shown that breaks in the power law are related to the line-of-sight thickness of galactic disks.  Observations of the LMC disk with IR emission by \citet{block10} show a two component power spectrum with a break at 100--200 pc, occurring at a depth that is comparable to the disk line-of-sight thickness.  The same break in the power spectra for the LMC at 100~pc is also seen with HI observations \citep{elmegreen01}, interpreted as the line-of-sight thickness.  If the line-of-sight thickness is responsible for the observed break in power spectra, dwarf galaxies could lack the break as the line-of-sight depth is comparable to the transverse length \citep{westerlund97,roychowdhury10}.  Increasing the scale height results in an increase of chance alignment along the line of sight, serving to decrease both the clustering scale length and magnitude of clustering.  

Interestingly enough, the youngest population ($\leq 10$~Myr) of star clusters observed by \citet{scheepmaker09} were observed to have the shallowest slope ($\alpha = -0.4$), where their oldest star clusters ($30 \leq \rm{Age (Myr)} < 400$) have the steepest slope of $\alpha = -0.8$.  Our youngest clusters on average (Class 3) display the most prominent clustering behavior whereas the oldest clusters on average (Class 1) systematically show minimal clustering with a very shallow slope.  Despite being fairly flat, the clustering results of our young clusters within Class 0 which have not been visually classified and could still include non-clusters, appear to be consistent with other observations, even if they fall short of the measured slopes of $\alpha \sim -0.8$ by \citet{zhang01}.  While these early results seem to suggest that the correlation function of young clusters may be similar between different galaxies (i.e., it hints toward a universal fractal dimension of hierarchical star formation), more studies of different galaxies are necessary to determine the validity of this statement.

As shown in Figure \ref{fig:2pcfage}, the strength of the clustering decreases with both increasing spatial scale and increasing age, excluding our multiple peak clusters (Class 3), though the increase in the clustering is only marginal for the symmetric clusters (Class 1).   At ages younger than 20 Myr, we see a transition from the stellar clusters displaying a highly clustered distribution toward a flat, non-clustered distribution for ages greater than 40 Myr.  This is consistent with star clusters born within a highly clustered structures and the clustering dissipating in a little as a few Myr after the formation from random motions of the clusters and shear effects.  

\begin{figure*}
\epsscale{1.2}
\plotone{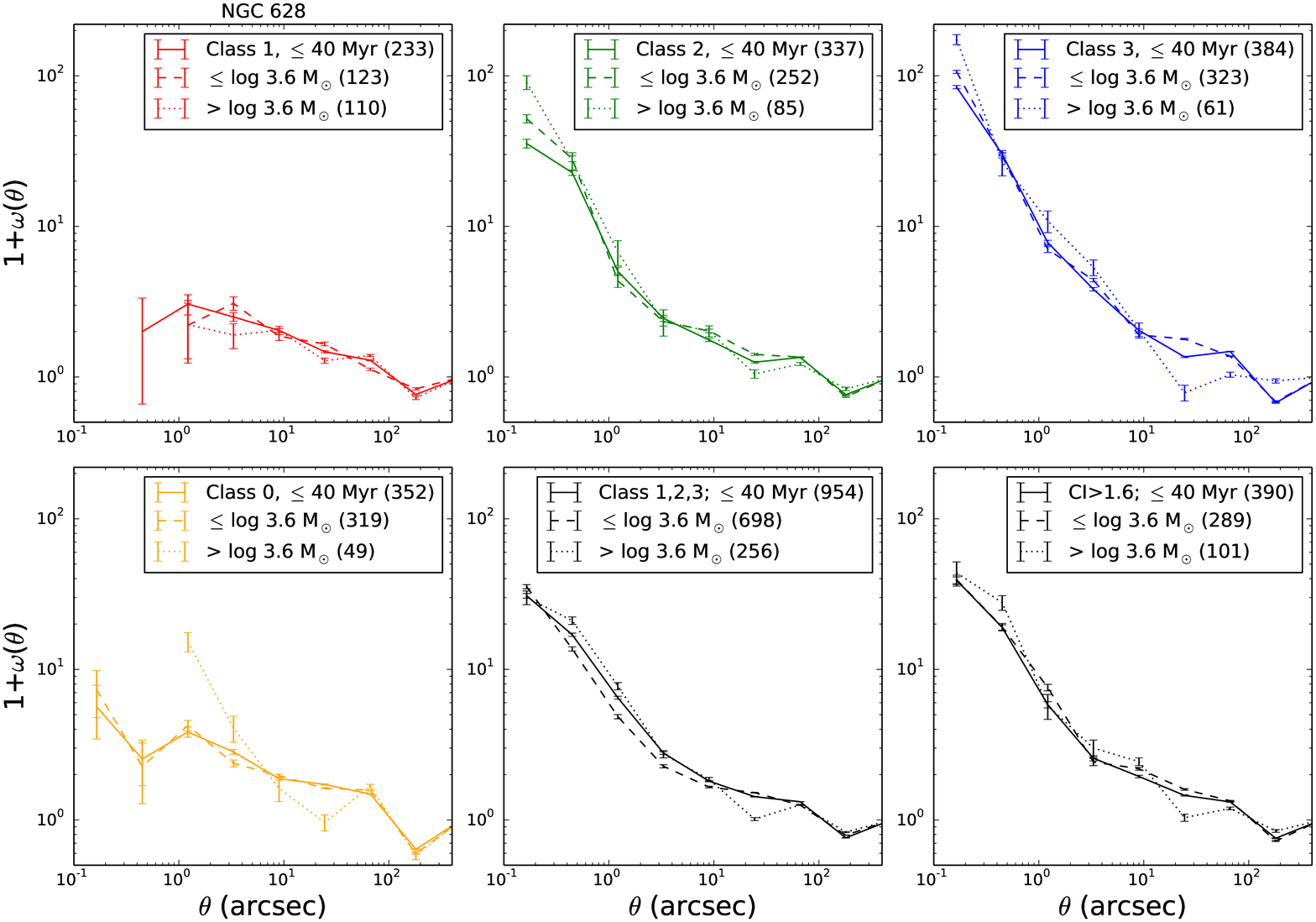}
\caption{
The two-point correlation function $1+\omega(\theta)$ as a function of angular distance (arcsec) for young ($<40$ Myr) clusters in NGC~628, divided into bins below (dashed line) and above (dotted line) the mass big of $\log(M/M_{\odot}) = 3.6$.  
\label{fig:hilomass}}
\end{figure*}

We see a slight increase in clustering with decreasing mass (Figure \ref{fig:2pcfmass}), predominatly only affecting the smallest spatial scales.  This is not an expected result as in a hierarchy/turbulence model universe, everything is self similar so there should be no mass dependency on the clustering results.  We know that there is an age effect on the clustering correlation and to understand if there is a bias present that makes our selection of low-mass clusters younger, we plot clusters younger than 40 Myr divided into high and low mass bins, shown in Figure \ref{fig:hilomass}.  When only considering our youngest clusters, the dependency of the mass of a cluster on the clustering strength disappears and we can see that the correlation is nearly identical for clusters above and below our mass bin of 3900 $M_{\odot}$.  Thus the mass dependence seen in Figure \ref{fig:2pcfmass} is probably an age effect (i.e., low mass clusters tend to be younger and more correlated than high mass clusters).   We lack number statistics in NGC 628 to be able to investigate how mass influences the clustering behavior for clusters older than 40 Myr and we defer it to future studies.  

We see that the opposite behavior occurs for the broadest Class 0 candidate cluster sources as there is a steep rise in the clustering for massive clusters (Figure \ref{fig:2pcfmass} and \ref{fig:hilomass}).  However, as we are dealing with small number statistics (there are only two objects within the first two bins of the CI $>$ 1.6 and mass $>$ 3900 $M_{\odot}$ bin) the clustering strength increase is accompanied with large error bars, making any conclusion unreliable at best.  With greater cluster numbers, we will be able to statistically investigate if the properties of the star clusters change as a function of environment and if this is reflected in a change of the observed clustering properties.

\section{Summary and Conclusion}\label{sec:summary}
In this paper, we present a technique to investigate the spatial clustering of the young stellar clusters in the LEGUS galaxy NGC~628 with UV+optical data taken with WFC3/UVIS from HST.  The inclusion of NUV observations within this study provides reliable measurements of both the age and masses of the clusters, giving us an unprecedented, high-angular resolution view of the clustering in a way that has not been possible before.  

Through visual inspection, we have identified 1264 stellar clusters within the galaxy that have an absolute V-band magnitude brighter than $-6$, CI $>1.4$, and detection in at least four filters.  For the fainter (V-band magnitude fainter than $-6$) clusters that did not undergo visual inspection (a total of 384 objects) that still have detection in at least four filters, we take cuts in both concentration index and age to separate out possible cluster candidates from spurious, non-cluster candidates.  Using this method, we identify 128 possible clusters of out a total of 384 faint sources, increasing the number of possible stellar clusters available for analysis to 1392.  

We implement the two-point correlation function to describe and study the amount of clustering occurring within all our cluster candidates.  We find that the clusters of NGC 628 have correlation functions that can be well described with a power law slope consistent with what is expected if the clustering is part of a hierarchy and that the observed clustering strength increases within younger clusters.  Our results reveal the internal hierarchical morphology present in the spatial distribution of the stellar clusters.  We find that the general observed hierarchy of the young stellar clusters decreases monotonically across the dynamical range of the galactic disk from 7 -- 10 kpc.  

Our clusters with centrally concentrated light profiles have the highest average mass and oldest ages and show the flattest power law slope of $\alpha \sim-0.14$ across all spatial scales.  Our multiple peaked clusters have the steepest slope of $\alpha \sim-1.51$ and correspond to the clusters with the lowest average masses and youngest ages in the galaxy.  The asymmetrical clusters display a slope of $\alpha \sim -0.82$.  The weighted average of the measured power law slope for the combination of all the genuine clusters is $\alpha \sim-0.8$.  These results from the correlation function reveal that our cluster population is part of a self-similar distribution with an average two-dimensional fractal distribution D2 = 1.2.  This fractal distribution is consistent with the hierarchical morphology of star forming structures formed from fractal gas that has undergone turbulent fragmentation.  

Both the asymmetrical and multiple peaked sources show a break in their power law, at 3\farcs3 and 1\farcs2 respectively, corresponding to a spatial scale length of 158~pc and 58~pc; the average slope at length-scales beyond the breakpoint is $\alpha \sim -0.24$.  These results suggest that the stellar clusters do form in clustered structures at the small spatial scales, while the clustering is erased and the spatial distribution is observed to be more homogeneous at larger spatial scales beyond the break at 3\farcs3.  Conversely, the symmetrical clusters have a single shallow slope of $\alpha\sim-0.14$, indicating nearly uniform distribution across all scales.  The measured slope for the faint non-visually classified sources have a power law slope of $\alpha \sim-0.20$; when we exclude sources with measured CI values less than 1.6, the slope steepens to $\alpha \sim-0.31$, which is still below what we find for the average of the genuine clusters in the survey.  The asymmetrical and multiple peak clusters dominate the observed clustering structure.  

In addition to the clustering strength decreasing with increasing spatial scale, we also find that the clustering strength decreases with increasing age, consistent with the fractal distribution being erased as the clusters age.  We find a dramatic decrease in the clustering at a timescale of 40~Myr.  We find that dividing up clusters by mass has a minor influence on the clustering behavior, however, the low-mass bin ($<4000$~M$_{\odot}$) is subject to stochastic IMF effects and as such, these masses are highly uncertain.  The mass dependency on the clustering behavior vanishes when we only consider the young clusters below 40~Myr. 

Future investigations of a larger sample of LEGUS galaxies will provide a better understanding of whether there is a common correlation length observed in the stellar clusters across galaxies and if the clustering depends on the ambient galactic environment.  The clustering nature of the young stellar clusters will not only provide a window to investigate whether or not star formation occurs in hierarchical patterns in both space and time, it will also inform us on the clustering properties of stars at the time of their formation, providing important information on the nature of the star formation process itself.  If stars and clusters map the same type of hierarchy, this would allow the use of observations on the clustering distribution of stellar clusters to trace and study star formation, providing a way to understand the cluster formation efficiency in relation to the star formation efficiency.  


\acknowledgements
We thank the anonymous referee for the careful
and constructive comments that greatly improved this
paper.

This paper is based on observations made with the NASA/ESA Hubble Space Telescope, obtained at the Space Telescope Science Institute, which is operated by the Association of Universities for Research in Astronomy, under NASA Contract NAS 5--26555.  These observations are associated with Program 13364 (LEGUS).  Support for Program 13364 was provided by NASA through a grant from the Space Telescope Science Institute.  

This research has made use of the NASA/IPAC Extragalactic Database (NED) which is operated by the Jet Propulsion Laboratory, California Institute of Technology, under contract with NASA.  

DAG acknowledges financial support by the German Research Foundation (DFG) through grant GO 1659/3-2.  MF acknowledges support by the Science and Technology Facilities Council [grant number ST/L00075X/1].

\appendix 
\section{Cluster Identification within Class 0}\label{sec:appA}
\begin{figure}
\epsscale{1.2}
\plotone{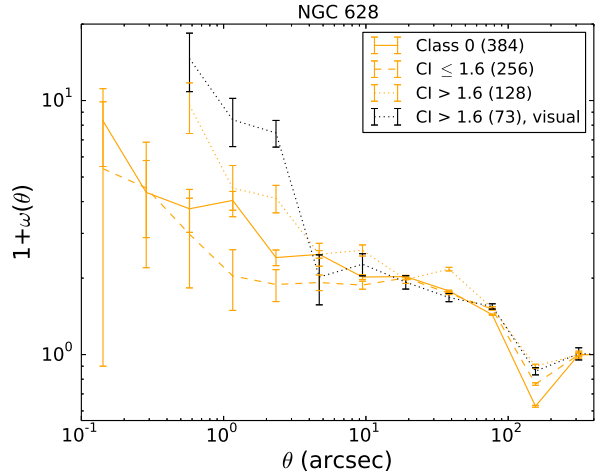}
\caption{
The correlation function for the non-visually identified Class 0 sources, divided into bins of high and low-CI (orange lines).  The black line shows the Class 0 sources that have been visually checked to verify that they are genuine clusters.  
\label{fig:2pcfclass0vis}}
\end{figure}

In order to ensure that our method selecting the broad (CI $>$ 1.6) sources within the Class 0 candidates accurately identifies genuine star clusters, we visually checked the 128 Class 0 sources with CI $>$ 1.6, finding that 73 are true star clusters (57\%).  While this is a high contamination rate and does affect our cluster results, placing a limit for sources at CI $>$ 1.6 does improve the blind selection of clusters and allows us to identify a correlation at a much higher degree of confidence than without the selection (dotted orange versus black in Figure \ref{fig:2pcfclass0vis}).  Additionally, small numbers limits our results of NGC 628 and we will only fully be able to investigate more significant results when we employ the galaxies within a larger selection of the LEGUS sample. 

\end{document}